\documentclass[sigconf]{acmart}
\PassOptionsToPackage{table}{xcolor}
\usepackage{tabularx}
\AtBeginDocument{%
  \providecommand\BibTeX{{%
    \normalfont B\kern-0.5em{\scshape i\kern-0.25em b}\kern-0.8em\TeX}}}

\copyrightyear{2025}
\acmYear{2025}
\setcopyright{cc}
\setcctype{by-nc}
\acmConference[CHI '25]{CHI Conference on Human Factors in Computing Systems}{April 26-May 1, 2025}{Yokohama, Japan}
\acmBooktitle{CHI Conference on Human Factors in Computing Systems (CHI '25), April 26-May 1, 2025, Yokohama, Japan}\acmDOI{10.1145/3706598.3714255}
\acmISBN{979-8-4007-1394-1/25/04}

\usepackage{tabularx}
\usepackage{booktabs}
\usepackage{multirow}
\usepackage{makecell}
\usepackage{colortbl}
\usepackage{booktabs}
\usepackage{array}
\usepackage{pgf}
\usepackage{collcell}
\usepackage{hhline}
\usepackage{longtable}
\usepackage{multirow}
\usepackage{amsmath}

\usepackage{wrapfig}
\usepackage{float}
\usepackage{enumitem}
\usepackage{url}
\usepackage{hyperref}
\usepackage{ragged2e}
\usepackage{tabularray}
\usepackage{setspace}
\usepackage{subcaption}
\usepackage{utfsym}

\usepackage{fontawesome}
\usepackage{xcolor}

\definecolor{high}{HTML}{628c77} 
\definecolor{mid}{HTML}{b9f2d5}  
\definecolor{low}{HTML}{FFFFFF}  

\definecolor{high1}{HTML}{7262ac} 
\definecolor{mid1}{HTML}{FFFFFF}  
\definecolor{low1}{HTML}{2e974e}  

\usepackage{graphicx}
\definecolor{darkred}{HTML}{7e0f12}
\definecolor{darkgreen}{rgb}{0.0, 0.5, 0.0}
\definecolor{purple}{HTML}{7262ac}
\begin{document}

\title{Deconstructing Depression Stigma: Integrating AI-driven Data Collection and Analysis with Causal Knowledge Graphs}



\author{Han Meng}
\email{han.meng@u.nus.edu}
\orcid{0009-0003-2318-3639}
\affiliation{
  \institution{Department of Computer Science, National University of Singapore}
  \streetaddress{21 Lower Kent Ridge Road}
  \country{Singapore}
  \postcode{119077}
}
\author{Renwen Zhang}
\email{r.zhang@nus.edu.sg}
\orcid{0000-0002-7636-9598}
\affiliation{
  \institution{Department of Communication and New Media, National University of Singapore}
  \streetaddress{21 Lower Kent Ridge Road}
  \country{Singapore}
  \postcode{119077}
}
\author{Ganyi Wang}
\email{ganyi-w@comp.nus.edu.sg}
\orcid{0009-0009-3294-6363}
\affiliation{
  \institution{School of Computing, National University of Singapore}
  \streetaddress{21 Lower Kent Ridge Road}
  \country{Singapore}
  \postcode{119077}
}
\author{Yitian Yang}
\email{yang.yitian@u.nus.edu}
\orcid{0009-0000-7530-2116}
\affiliation{
  \institution{Department of Computer Science, National University of Singapore}
  \streetaddress{21 Lower Kent Ridge Road}
  \country{Singapore}
  \postcode{119077}
}
\author{Peinuan Qin}
\email{e1322754@u.nus.edu}
\orcid{0000-0002-8737-8369}
\affiliation{
  \institution{Department of Computer Science, National University of Singapore}
  \streetaddress{21 Lower Kent Ridge Road}
  \country{Singapore}
  \postcode{119077}
}
\author{Jungup Lee}
\email{swklj@nus.edu.sg}
\orcid{0000-0002-8243-0543}
\affiliation{
  \institution{Department of Social Work, National University of Singapore}
  \streetaddress{21 Lower Kent Ridge Road}
  \country{Singapore}
  \postcode{119077}
}
\author{Yi-Chieh Lee}
\email{yclee@nus.edu.sg}
\orcid{0000-0002-5484-6066}
\affiliation{
  \institution{Department of Computer Science, National University of Singapore}
  \streetaddress{21 Lower Kent Ridge Road}
  \country{Singapore}
  \postcode{119077}
}


\begin{abstract}
Mental-illness stigma is a persistent social problem, hampering both treatment-seeking and recovery. 
Accordingly, there is a pressing need to understand it more clearly, but analyzing the relevant data is highly labor-intensive. 
Therefore, we designed a chatbot to engage participants in conversations; coded those conversations qualitatively with AI assistance; and, based on those coding results, built causal knowledge graphs to decode stigma. 
The results we obtained from 1,002 participants demonstrate that conversation with our chatbot can elicit rich information about people’s attitudes toward depression, while our AI-assisted coding was strongly consistent with human-expert coding.
Our novel approach combining large language models (LLMs) and causal knowledge graphs uncovered patterns in individual responses and illustrated the interrelationships of psychological constructs in the dataset as a whole. 
The paper also discusses these findings’ implications for HCI researchers in developing digital interventions, decomposing human psychological constructs, and fostering inclusive attitudes.
\end{abstract}


\maketitle

\section{Introduction}

Mental illnesses are pervasive, with depression alone touching the lives of approximately 280 million people worldwide \cite{WHO_depression_2023}.
Nevertheless, due to \textit{mental-illness stigma} \cite{intervention_corrigan_1999}, many people with psychiatric disorders face social rejection, employment and housing discrimination, negative self-perceptions, and reluctance to seek help \cite{depression_stigma_peluso_2009, seek_help_reluctant_clement_2015}; and this stigma remains persistent despite efforts to eradicate it by fostering understanding and acceptance \cite{intervention_corrigan_1999}. 
The intricate nature of stigma, with its multifaceted manifestations and deep-rooted societal influences, makes it challenging to measure and eliminate \cite{attribution_theory_corrigan_2000}.

Existing methods for dissecting mental-illness stigma have various limitations. 
Self-report questionnaires \cite{attribution_model_corrigan_2003} provide relatively little detail, particularly about the hidden underpinnings of the respondents' emotions and actions \cite{comparison_taherdoost_2022}, and are prone to social-desirability bias \cite{sd_scale_inaccurate_furnham_1986}.
Social-media analysis has also been widely used to assess stigmatizing effects and cognitive distortions \cite{detect_method_fang_2023, detect_method_mittal_2023}, leveraging the large volume of diverse language samples available online. 
However, discourse on stigma-related topics in such posts tends to be brief and shallow \cite{social_media_decontext_boyd_2012}, and this type of analysis suffers from population bias \cite{social_media_deidentify_ruths_2014}.

A promising solution is the collection and analysis of in-depth interview data, which contains rich nuances of how mental health is discussed, perceived, and construed \cite{interview_stigma_measure_liggins_2005, interview_stigma_measure_lyons_1995}.
Interviews, a form of conversational interaction, serve as space and medium for unfolding perspectives, facilitating social-knowledge exchanges, and reproducing social norms \cite{conversation_importance_jenlink_2005, sensitive_kvale_2009}.
Such conversational contexts, along with interviewing techniques like probing questions and reflective listening \cite{sensitive_kvale_2009}, allow interviewers to identify recurring patterns such as the use of derogatory terms, casual jokes, or dismissive comments about mental illness that reveal stigma, whether intentional and/or unintentional \cite{conversation_importance_meredith_2019, microaggression_stigma_gonzales_2015}. 
However, gathering and dissecting interview data in quantities large enough to attain such insights has generally been infeasible due to the enormous amounts of time and effort required \cite{coding_manual_saldana_2016}, as well as the sensitive nature of mental illness-related discussions.

The burgeoning field of AI could help to reduce those burdens. 
AI-powered chatbots have already been used to simulate human interviewers to gather qualitative data on sensitive topics \cite{chatbot_reduce_kim_2020, chatbot_reduce_sebastian_2017}, which can reduce social-desirability bias and foster self-disclosure \cite{chatbot_disclosure_lucas_2014}; and some computational techniques, such as machine learning \cite{detect_method_jilka_2022} and word embedding \cite{detect_method_mittal_2023}, have been applied to automate stigma detection in large textual datasets. 
These advancements have facilitated the development of fruitful AI methods that excel at rapidly and accurately coding qualitative data like interviews for large-scale analysis \cite{ai_qualitative_feuston_2021}. 
The advent of LLMs could progress this capability by enabling real-time identification of language subtleties, emotional tones, and behavioral indicators that may not be immediately apparent to human coders \cite{coding_deductive_llm_tai_2024}.
However, AI/LLMs rely on syntactic and semantic cues, and their grounding in psychological theories/models ranges from simplistic to nonexistent \cite{theory_nlp_boyd_2021}. 
This would likely lead them to focus on end-state classification rather than on how, when, or why discriminatory behavior develops.

These limitations of AI/LLM-assisted qualitative-data analysis can be partly addressed by incorporating it with causal knowledge graphs (CKGs), which may be more capable of unraveling the mechanisms, processes, and triggers of stigmatizing behavioral intention. 
CKGs model causation via well-structured representations of entity-relationship-entity triples \cite{causal_graph_overview_jaimini_2022}. 
We consider that, in alignment with appropriate theoretical frameworks, CKGs could enable the mapping of stigma mechanisms onto established factor-pathway models and illustrate the interrelationships among various stigma-related constructs. 
Synergizing LLM capabilities with CKGs would aggregate the inductive strengths of LLMs, i.e., their ability to discover latent relationships, with the deductive power of CKGs: in this case, psychological-hypothesis generation \cite{ckg_llm_tong_2024}.

In this study, we examine how AI/LLM-assisted methods and CKGs can be integrated to both collect and analyze data related to mental-illness stigma, with a particular focus on the stigma attached to \textit{depression} due to its global prevalence. Specifically, we ask the following research questions: 



\begin{quote}
\textbf{RQ1.} \textit{To what extent can AI-assisted qualitative-data collection and analysis methods effectively capture depression stigma?}



\textbf{RQ2.} \textit{How well can the integration of CKG and LLM reveal the constructs that drive depression stigma and their interrelationships?}
\end{quote}

Accordingly, the present study's methodology began with using an AI chatbot as an interviewer to collect qualitative data about perceptions of depression from 1,002 participants.
We then applied AI-assisted qualitative coding to identify stigma attributions, guided by a codebook developed through human coding. 
To validate this workflow, we compared our AI-assisted coding with human-generated codes and existing computational methods. 
We then semi-automatically constructed a CKG that could portray the interplay of stigma, dissect psychological constructs across conversations, discern similarities and differences in the participants' logic chains, and intertwine these elements to build a conceptual model. 
Our integration of LLM and CKG techniques involved the former suggesting new relationships and the latter providing a structured framework for deducing claims and hypotheses. 
This dual method uncovered constructs like \textit{personalities} and \textit{past experiences} that are supported by existing theories \cite{personality_reviewer_steiger_2022, experience_weinstein_1989}, along with new causal relationships such as the impact of \textit{personalities} on \textit{emotional responses} and \textit{situations} on \textit{anticipated discriminatory behaviors}.

Our work makes several contributions to the HCI community. 
First, it introduces an \textbf{AI-assisted data collection and analysis pipeline} that leverages AI to systematically elicit and analyze latent psychological constructs, such as depression stigma.
This method offers a non-labor-intensive yet nuanced approach to capturing large volumes of interview data and understanding and coding attributions of depression stigma.
Second, its \textbf{integration of CKGs with LLMs} facilitates interpretation of the causal networks between various psychological constructs, such as cognition, emotions, and behavioral intentions related to mental health. 
This contributed to the development of a conceptual model that illuminates the predictors and processes of stigma formation.
And third, this study's findings can inform the design of \textbf{tailored, theoretically grounded micro-interventions} to combat depression stigma by enabling real-time stigma identification and the creation of CKGs specific to individuals.
This personalized approach may address the limitations of one-size-fits-all interventions, enhancing effectiveness while reducing potential unintended consequences.






\section{Related Work}

\subsection{AI Applications in Mental Health}

AI has been increasingly integrated into digital mental-health care delivery \cite{ai_mh_ma_2023, ai_mh_ma_2024}, offering diverse applications spanning psychotherapy \cite{therapy_ai_prochaska_2021}, psychoeducation \cite{ai_psychoedu_jang_2021}, and social companionship \cite{ai_mh_ma_2023}.
Modern AI systems have demonstrated remarkable capabilities in psychiatric care through their ability to engage in naturalistic, human-like therapeutic interactions \cite{ai_mh_jo_2023}.
Functioning as both clinical tools and digital companions, these AI systems offer multiple benefits, including empathic communication, non-judgmental responses, regular check-ins, and tailored feedback \cite{ai_mh_li_2023}.

However, despite the promising potential of AI to promote mental well-being through improved accessibility and diverse support mechanisms, \textit{social stigma} remains a primary barrier that may prevent people from seeking help through these digital systems, just as it has historically deterred people from accessing traditional mental health services \cite{ai_mh_hoffman_2024}.
This persistent challenge of stigma therefore demands careful attention as we continue to develop and design AI-based mental-health interventions.

\subsection{Mental-illness Stigma}

Social stigma, as originally defined by Goffman, includes regarding mental illness as divergent from what society considers normal and correct, and mentally ill individuals as tainted and devalued \cite{stigma_spoiled_identity_goffman_1964}. 
It stems from stereotypes and prejudices \cite{attribution_model_corrigan_2003} that frequently manifest as unconscious bias, i.e., negative attitudes or cognition that can sway people's decisions without their awareness \cite{microaggression_stigma_gonzales_2015}. 
Social stigmatization of people with mental illness is very prevalent worldwide \cite{measuing_stigma_corrigan_2010}. 
The pervasiveness of public stigma impedes help-seeking, treatment-seeking, and recovery while also exacerbating historical injustices \cite{decolonial_pendse_2022} and creating obstacles to employment, housing, and social connections \cite{understanding_corrigan_2002}. 
Consequently, only about 50\% of people with depression, for example, seek treatment \cite{depression_ratio_kessler_2003}, and many discontinue it to avoid being labeled psychiatrically \cite{depression_seeking_dew_1988}. 
Understanding and reducing mental-illness stigma is therefore crucial to social welfare and the promotion of inclusive attitudes.

The \textit{attribution model} \cite{attribution_model_corrigan_2003} is a theoretical framework that has guided many explorations of the factors contributing to mental-illness stigma. 
It includes three dimensions – \textit{personal responsibility}, \textit{emotional responses}, and \textit{behavioral responses} – and holds that individuals' perceptions of the extent to which a person with a mental illness is responsible for their condition can lead to stigmatizing emotional responses, such as anger, fear, and lack of pity. 
Such responses may then elicit discriminatory behaviors including coercive segregation, social distance, and the withdrawal of help. 
The attribution model serves as a solid foundation for stigma measurements and forms the theoretical basis for our study's deconstruction of stigma.

Virtual reality \cite{realtime_martinez_2024}, videogames \cite{personalized_anvari_2024}, and social-media campaigns \cite{reduce_sm_feuston_2019} have all shown promise for reducing mental-illness stigma by raising awareness and/or facilitating positive social contact. 
Chatbots have been effective at delivering such interventions, in part because they offer anonymity and encourage self-disclosure \cite{chatbot_reduce_kim_2020, disclosure_lee_2022, chatbot_aq27_practice_lee_2023}. 
However, many existing anti-stigma approaches adopt a one-size-fits-all strategy that is likely to ignore how stigma manifests differently across social contexts, potentially limiting their efficacy or even causing them to backfire \cite{backfire_dobson_2022}.

\subsection{Measuring and Deciphering Mental-illness Stigma}

To develop any successful anti-stigma initiative, one must identify the factors and mechanisms that create and sustain the targeted stigma \cite{vignette_link_1987}.
This section reviews three approaches to deciphering mental-illness stigma: traditional quantitative methods, computational techniques, and traditional qualitative analysis.

\subsubsection{Traditional Quantitative Methods} 

Quantitative-scale protocols have been widely used to assess various components of mental-illness stigma, including behavior \cite{vignette_link_1987}, stereotyping \cite{stereotyping_bedini_2000}, cognitive separation \cite{cognitive_ostma_2002}, emotional reactions \cite{affective_vezzoli_2001}, and status loss and discrimination \cite{status_loss_secker_1999}. 
One notable instrument is the \textit{Attributional Questionnaire} \cite{attribution_model_corrigan_2003}, a causality tool that helps to unravel the genesis and perpetuation of stigma by assessing the key constructs defined in Corrigan et al.'s \cite{attribution_theory_corrigan_2000} social-cognitive models.
These tools have greatly facilitated the understanding of mental-illness stigma.

Despite the widespread use of standardized self-report surveys to measure various constituents of mental-illness stigma, such instruments might under-represent the complex lived experience of stigmatizing and being stigmatized, the mechanisms of stigma's progression, or other societal aspects of these processes \cite{interview_stigma_measure_liggins_2005}. 
Moreover, this measurement approach is susceptible to social-desirability bias \cite{vignette_link_1987}. 
That is, anti-stigma campaigns \cite{campaign_corrigan_2001} and other public-education efforts \cite{intervention_corrigan_1999} have emphasized the importance of not rejecting individuals simply because they have sought mental-health treatment. 
As a result, even if they privately hold stigmatizing attitudes, people are unlikely to express them openly, because they want to appear enlightened and caring \cite{sd_scale_inaccurate_van_2008}. 
This dissimulation can lead scholars to underestimate true stigma levels and/or to misidentify genuine predictors of stigma \cite{sd_stigma_michaels_2013}. 
Thus, researchers should employ multiple approaches to decoding stigma \cite{vignette_link_2004}, develop measures that mitigate potential confounders, and explore modes of interpretation that better capture the intricacies of stigma.

\subsubsection{Computational Techniques}

Social-media platforms like Twitter and Weibo include masses of largely unfiltered qualitative data relevant to mental health \cite{social_media_bail_2017, social_media_pavlova_2020}. 
Leveraging these resources, some researchers have employed various computational techniques to analyze mental-illness stigma, studying conditions such as depression \cite{detect_method_li_2018}, schizophrenia \cite{detect_method_jilka_2022}, and substance use \cite{detect_method_roesler_2024}.
Specifically, Fang \& Zhu used linguistic-analysis tools such as LIWC \cite{detect_method_fang_2023}, while Mittal et al. developed frameworks that assess stigmatizing discourse in social media and news content by creating custom dictionaries and calculating word-embedding similarities \cite{detect_method_mittal_2023}.

Machine-learning (ML) and deep-learning (DL) solutions have also contributed to this field, by enabling real-time, automated mental-illness stigma detection and classification within large social-media datasets \cite{detect_method_robinson_2019}.
These approaches range from binary 'present/ absent' classification of such stigma using algorithms such as SVM, Random Forest \cite{detect_method_jilka_2022}, and BERT \cite{detect_method_lee_2022} to more sophisticated models that use $n$-grams to differentiate among types of stigma \cite{detect_method_li_2018} or types of stigma-related language, e.g., metaphors, personal stories, ridicule, and jokes \cite{detect_method_roesler_2024}.

ML and DL methods can efficiently and rapidly classify large volumes of social-media content pertinent to stigma. 
However, social-media posts provide relatively brief snippets of stigma-related discussions, and their decontextualized aggregation into datasets may fail to capture the full panorama of stigma-related interactions and their evolving dynamics \cite{social_media_decontext_boyd_2012}. 
Additionally, social-media data suffer from population biases \cite{social_media_deidentify_ruths_2014} that vary across platforms, potentially skewing interpretation. 
More importantly, these approaches mainly focus on classifying and/or quantifying words and phrases, and thus are likely to oversimplify the psychological underpinnings of stigmatizing behavior while largely ignoring its processes and triggers. 
Given these drawbacks, there is a pressing need for a more process-oriented analysis of stigma toward people suffering from mental illnesses in more context-rich settings.

\subsubsection{Traditional Qualitative Analysis}

Human narratives are rife with psychological constructs, perceptions, and reasoning about societal topics, and act as catalysts for deep reflection and genuine free exchange of ideas \cite{conversation_importance_jenlink_2005}. 
Interview data contain valuable information about how social problems are perpetuated, unfolded, or negotiated \cite{conversation_importance_jenlink_2005}, reflecting the endogenous organization of social activities \cite{conversation_banathy_2005}. 
Such conversations naturally create a space and medium for inclusion, connection, and self-disclosure, allowing interviewers to uncover the underlying beliefs, attitudes, and experiences that shape people's perspectives and may lead to evolving consciousness \cite{conversation_bohm_2004, conversation_banathy_2005}.

Specifically, interviews can capture subjective experience/un-derstanding of their focal phenomenon and the complex social systems that give rise to it \cite{interview_stigma_measure_liggins_2005, interview_stigma_measure_lyons_1995}.
Like participant observation, interviewing can thus provide insiders' views of stigma's dynamics and impacts that rating scales or analysis of social-media data could overlook. 
For instance, Lyons and Ziviani \cite{interview_stigma_measure_lyons_1995} effectively used interviews and participant observation to track changes in stereotypical beliefs and preconceptions about the mentally ill. 
Liggins and Hatcher \cite{interview_stigma_measure_liggins_2005} likewise applied interviews to reveal how people express fear and hopelessness toward mental illness, and showcased the depth of the insights that qualitative analysis makes possible.

Although data from interviews can facilitate a deep understanding of social phenomena, they also pose major challenges, as collecting, coding, categorizing, and interpreting them requires substantial time, effort, and expertise \cite{human_coding_bias_leeson_2019} as well as strong theoretical knowledge \cite{coding_manual_saldana_2016}. 
AI's eloquence in processing natural language makes it adept at collecting interview data \cite{chatbot_kim_2019}, and its perceptiveness may enable it to streamline the analysis process while preserving the fertility of insights \cite{deductive_labelling_xiao_2023}. 
Yet, explorations of their ability to profile stigma based on interview data remain rare.

\subsection{AI-assisted Qualitative-data Collection and Analysis} 

Chatbots \cite{chatbot_hoque_2024} offer a promising approach to interview-data collection for exploratory research. 
They can provide a middle ground between questionnaires and traditional in-person interviews, offering a more engaging experience \cite{chatbot_kim_2019} than the former while being more scalable than the latter. 
In particular, AI-powered chatbots are flexible, fast, and adaptable tools that can gather rich qualitative insights into complex social and psychological phenomena by combining gregarious interactivity with wide reach \cite{chatbot_xiao_2020}. 
Moreover, AI can create materials tailored to fostering chatbot users' self-disclosure \cite{disclosure_lee_2022}.

While these data-collection methods provide rich qualitative information, deciphering and interpreting this wealth of data is equally crucial to unlocking its full potential. 
AI technologies are also revolutionizing qualitative-coding processes \cite{ai_qualitative_feuston_2021, ai_qualitative_muller_2016, ai_qualitative_rietz_2020}, mitigating manual methods' time and resource constraints. 
In deductive coding, AI has applied predefined codebooks to large datasets \cite{coding_deductive_llm_tai_2024, deductive_labelling_xiao_2023}, and achieved fair to substantial agreement with human-expert coders \cite{cohens_kappa_mchugh_2012, prompt_practice_dunivin_2024}. 
In inductive coding, meanwhile, AI can help researchers initialize codes, uncover themes, and synthesize essences from raw data \cite{content_analysis_coding_toolgpt_gao_2023}. 
Together, these AI-assisted approaches not only enhance efficiency, but also pave the way to the discovery of new constructs, patterns, and conceptualizations. 






\subsection{Synergy between Large Language Models and Causal Knowledge Graphs}

Importantly, however, the complex interplay of contexts, relational dynamics, and reasoning chains within qualitative data – which is particularly crucial to understand when examining intricate societal phenomena – may extend beyond what any single descriptive label can convey. 
This points to the need for additional, integrative approaches that are both analytical and explanatory.

CKGs represent a powerful fusion of structured-information representation and intelligent reasoning, and can capture both overt and covert relationships between entities in a domain \cite{kg_carta_2023}.
Indeed, they excel at integrating information from diverse sources, uncovering hidden patterns in it, and facilitating efficient navigation of complex knowledge landscapes \cite{ckg_uleman_2021, ckg_borsboom_2021}. 
Complementing CKGs, LLMs can access and articulate domain knowledge previously confined to human experts \cite{ckg_llm_kiciman_2024} and excel at lexical, syntactic, and semantic analysis and conceptual understanding.
The synergy between CKGs' structural prowess and LLMs' advanced linguistic capabilities \cite{ckg_llm_tong_2024} allows researchers to map out and explore intricate relationship networks \cite{kg_llm_pan_2024}. 
This synergistic approach echoes the holistic- vs. analytic-cognition dichotomy in social psychology \cite{human_psych_nisbett_2001}, and has been validated by prior studies showing its effectiveness in simulating bio-psycho-social interactions \cite{ckg_borsboom_2021, kg_psych_crielaard_2022}.
Such integration represents a promising new frontier in HCI and psychological research: not only enhancing researchers' ability to deduce relationships between factors, but also guiding the generation of latent novel causal hypotheses \cite{ckg_llm_tong_2024}.

Having recognized the potential of interview data to elicit disclosure and promote self-reflection, the capabilities of LLMs to imitate interviewers and interpret semantic and linguistic cues, and the power of CKGs to represent intricate interrelationships, our aim is to leverage those elements synergistically to dissect stigma in an unprecedented way. 



\section{Methodology}
\label{sec:method}

\begin{figure*}
    \centering
    \includegraphics[width=0.9\linewidth]{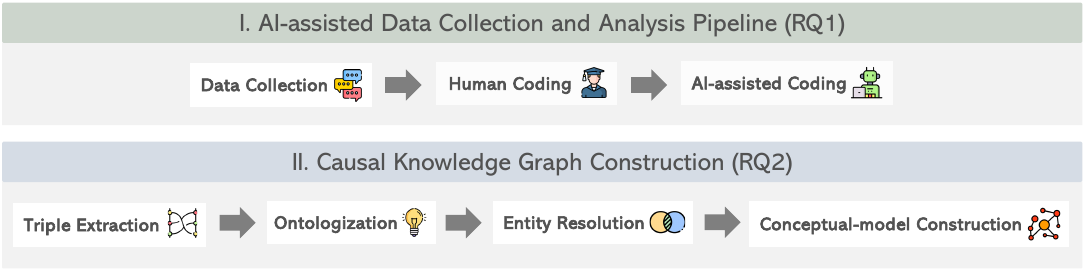}
    \caption{Methodology overview. In this work, we propose this approach to deconstruct depression stigma through two main phases: \textit{I. AI-assisted Data Collection and Analysis Pipeline} (RQ1) and \textit{II. Causal Knowledge Graph Construction} (RQ2).}
    \Description{Flowchart divided into two main sections. Section I 'AI-assisted Data Collection and Analysis Pipeline (RQ1)' shows a linear sequence of three steps: 'Data Collection,' followed by 'Human Coding', followed by 'AI-assisted Coding.' Section II 'Causal Knowledge Graph Construction (RQ2)' shows a linear sequence of four steps: 'Triple Extraction,' 'Ontologization,' 'Entity Resolution,' and 'Conceptual-model Construction.' All steps are connected by arrows showing the progression of the methodology.}
    \label{fig:overview}
\end{figure*}

To address the limitations of existing methods, including the lack of depth captured by questionnaire instruments \cite{comparison_taherdoost_2022}, the decontextualized nature of social-media posts \cite{social_media_decontext_boyd_2012}, and the burden of manual coding of interview transcripts \cite{coding_manual_saldana_2016}, we propose a novel approach to automatically unraveling depression stigma - a common and representative form of mental-illness stigma - by integrating chatbot interviews (Section \ref{method:datacollection}), AI-assisted qualitative coding (Section \ref{method:auto}), and CKG construction (Section \ref{method:ckg}) (Figure \ref{fig:overview}).

\subsection{AI-assisted Data Collection and Analysis Pipeline}

\begin{figure*}
    \centering
    \includegraphics[width=0.95\linewidth]{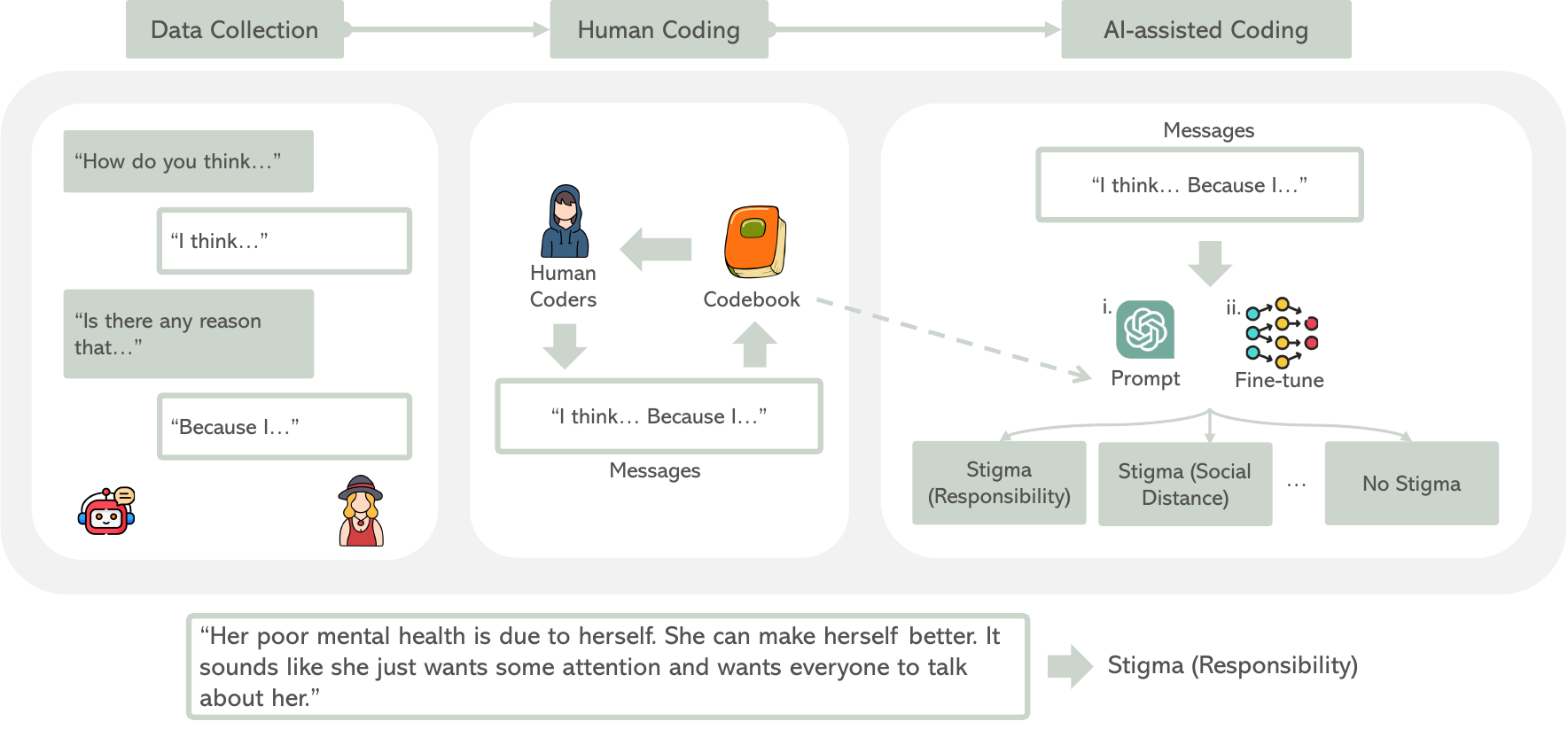}
    \caption{Overview of the AI-assisted data collection and analysis pipeline. This pipeline encompasses three main steps: \textit{Data Collection} to gather interview data using an AI-powered chatbot, \textit{Human Coding} to establish expert codes and develop a codebook, and \textit{AI-assisted Coding} to expand coding to larger datasets and detect and categorize stigma-related expressions.}
    \Description{Flowchart diagram organized into three sequential stages to illustrate the AI-assisted data collection and analysis pipeline. The diagram shows three major steps: 'Data Collection,' followed by 'Human Coding,' followed by 'AI-assisted Coding.' The Data Collection section illustrates human-chatbot interaction with dialog boxes. The 'Human Coding' section shows coders interacting with a codebook and participant messages. The 'AI-assisted Coding' section demonstrates the primary prompt-based approach to message classification, with fine-tuning methods shown as an alternative for comparison. The process is illustrated using a sample message about mental health, which is classified as 'Stigma (Responsibility).' All steps are connected by arrows showing the progression of the methodology.}
    \label{fig:individual}
\end{figure*}

Our three-step workflow (Figure \ref{fig:individual}) combined data collection using AI-assisted interviews, human coding to develop a codebook and expert codes, and AI-assisted coding to automate the analysis to identify attributions of depression stigma.
Each of these three phases is described in turn below.

\subsubsection{Data Collection}
\label{method:datacollection}

\paragraph{\textcolor{darkred}{\textbf{Procedure.}}}
Data collection centered on the participants' respective 20-minute interviews with an AI-powered chatbot, for which they were compensated US\$6.30. 
After obtaining their informed consent, we warned the participants that the upcoming interview scenario related to depression, and reiterated that they had the option to withdraw if this made them uncomfortable. 
Next, they provided their demographic information before proceeding to the main phase of the study, an interview with a chatbot we named \textit{Nova}. 
Its interface can be seen in the \textit{Supplementary Materials}.

Each chatbot interview started with a rapport-building \textit{small-talk} session, followed by Nova's delivery of a depression-related \textit{vignette}, during which it intermittently asked the participant for brief responses. 
After the vignette, Nova posed \textit{open-ended questions} prompting participants to share their opinions and related experiences. 
These core questions were split into two parts by a second brief session of small talk, aimed at re-engaging the participants with lighthearted discussion and relieving their potential emotional burdens. 

After the interaction with Nova ended, all participants were debriefed about common types of stigma and the study's objectives. 
All materials that emanated from the chatbot were refined by a mental-health specialist (a co-author) and an external consulting psychiatrist.

\paragraph{\textcolor{darkred}{\textbf{Participants.}}}
We amassed participants through two online research platforms, Prolific and Qualtrics. 
Our inclusion criteria were that they 1) were at least 21 years old, 2) spoke English as their first language, 3) were willing to read material related to mental illness, and 4) had no immediate or urgent mental-health concerns. 
The last criterion is incorporated because of the potential risk that vignettes about depression could cause distress and/or trauma to participants grappling with such issues \cite{ethic_mental_illness_roberts_2002}.
During recruitment, we clearly outlined our research's scope and duration, and the participants' right to withdraw from it at any time. 
We used IP addresses to filter out potential duplicate participants. 

We recruited 1,002 participants from Western countries, mostly the United States and the United Kingdom, with an average age of 46.4 ($SD = 16.45$). 
They were 53.9\% male, 45.9\% female, and 0.2\% other genders. 
Ethnically, 69.4\% were white, 21.0\% black, 4.7\% Asian, and 6.0\% other ethnicities. 
None of the participants reported having an ongoing mental illness, but 52.5\% said they had close friends or family members with mental-health issues.

\paragraph{\textcolor{darkred}{\textbf{Vignettes.}}}

\textit{Vignettes}, concise fictional narratives, are valuable research tools for exploring attitudes toward mental health \cite{vignette_alem_1999}. 
These stories, grounded in research and real experiences \cite{chatbot_aq27_practice_lee_2023, vignette_alem_1999}, allow research participants to respond to concrete scenarios through various feedback methods, and thus to provide researchers with insights into their perspectives. 
We created a vignette focused on "\textit{Avery}," a person of unspecified age and gender who was experiencing depressive symptoms. 
These symptoms were as set forth in the DSM-5 \cite{dsm5_apa_2013}, except that the more extreme ones such as self-harm and suicide were avoided, and no technical or medical jargon was used. 
The vignette, refined with expert input from our mental-health specialist and consulting psychiatrist, illustrated how such symptoms negatively impacted various aspects of Avery's life, including study, work, relationships, and interactions with others.

Specifically, all participants read the following vignette:

\begin{quote}
    \textit{Avery is employed by a company, and in their spare time, they are dedicated to lifelong learning, doing extensive reading and writing. However, Avery has been diagnosed with depression recently. It has become challenging for them to concentrate, resulting in a decline in work performance and learning outcomes. Interests that once brought them joy have waned. Avery has distanced themself from friends, becoming easily irritated in social interactions with colleagues and feeling judged by others. Avery lives with family and cannot do much, especially household chores. Social media intensifies their feelings of loneliness and anger, leading to frustration about the source of the anger.}
\end{quote}

\paragraph{\textcolor{darkred}{\textbf{Interview Questions.}}}

To elicit the participants' opinions, we designed multiple questions adapted from Corrigan et al.'s attribution model \cite{attribution_model_corrigan_2003} and the Attribution Questionnaire-27 (AQ-27) \cite{aq27_corrigan_2012}, which Nova posed as soon as they had finished reading the vignette. 
The AQ-27, developed from the attribution model, operationalized theoretical constructs into nine measurable items (i.e., \textit{blame}, \textit{fear}, \textit{pity}, \textit{anger}, \textit{helping}, \textit{avoidance}, \textit{coercion}, \textit{segregation}, and \textit{dangerousness}), each assessed through three standardized survey questions.
Following Lee et al. \cite{chatbot_aq27_practice_lee_2023}, we combined the coercion-segregation and fear-dangerousness pairs to create a more concise interview flow and adapted the survey questions into seven chatbot questions corresponding to the attributions of \textit{responsibility} (i.e., \textit{blame}), \textit{fear}, \textit{pity}, \textit{anger}, \textit{helping}, \textit{social distance} (i.e., \textit{avoidance}), and \textit{coercive segregation} (Table \ref{tab:question}).

We randomized the order of the questions to avoid priming effects \cite{priming_effect_molden_2014} and embedded each question within a vivid, relatable scenario to mitigate social-desirability bias and foster more honest responses \cite{sd_indirect_q_fisher_1993}.
All question selections and adaptations were validated by the mental-health specialist and consulting psychiatrist.

\paragraph{\textcolor{darkred}{\textbf{AI-assisted Data-collection Strategies.}}}

To facilitate self-disclosure, we employed three techniques. 
The first is that we formulated one to two \textit{follow-up questions} \cite{follow_up_q_han_2021} for each interview question to probe for underlying reasons, anticipated outcomes, or specific triggering scenarios.
Specifically, when dealing with emotional responses (i.e., \textit{anger}, \textit{fear}, and \textit{pity}) and \textit{responsibility} attributions, Nova prompted the user to give \textit{reasons} for their response if it fell below a minimum-length threshold. 
For behavior-related responses (i.e., \textit{coercive segregation}, \textit{social distance}, and \textit{helping}), it asked for reasons if the response was non-stigmatizing, and for \textit{potential results} if it was stigmatizing. 
For example, if a participant endorsed \textit{coercive segregation}, Nova might ask, \textit{"What do you think would happen if Avery was not involuntarily admitted to a psychiatric hospital?"} 

The second technique was \textit{active listening} \cite{active_strategies_zheng_2023}: i.e., Nova restated and validated the user's viewpoint and expressed understanding of it. 
The third and final technique was \textit{neutral self-disclosure}, in which Nova shared both positive and negative perspectives to elicit participants' disclosure \cite{disclosure_lee_2022} without shifting their attitudes (as further illustrated in the \textit{Supplementary Materials}). 

\paragraph{\textcolor{darkred}{\textbf{System Implementation.}}}

To create our AI-powered chatbot, we used the UChat platform and adopted a hybrid approach that combined pre-written scripts with AI-generated text. 
The small-talk questions, vignette delivery, questions based on the attribution model \cite{attribution_model_corrigan_2003}, and chatbot's neutral self-disclosure for each question were all pre-scripted and hard-coded. 
All other conversational elements, including follow-up questions and active responses to participants, were generated by the AI. 
We used \texttt{GPT-4-1106-preview} \cite{gpt4_achiam_2023} for generating the chatbot's text, with a maximum token limit of 100 and a temperature setting of 0.2. 
The chatbot interface was integrated into Qualtrics programming, and we ensured that the concurrency rate stayed below 50.


\begin{table*}[tbp]
\small
\caption{Chatbot question scripts and their corresponding AQ-27 survey items.}
\Description{Three-column comparison table showing the relationship between chatbot questions and a standardized questionnaire. The columns are labeled 'Chatbot Question Scripts,' 'AQ-27 Survey Item,' and 'Attribution Types.' Seven rows cover different aspects of depression stigma, with each row mapping a chatbot question to its corresponding survey item and attribution type.}
\label{tab:question}
\renewcommand{\arraystretch}{1.2}
\begin{tabular}{p{0.45\textwidth}p{0.3\textwidth}p{0.1\textwidth}}
\toprule
\multicolumn{1}{c}{\textbf{Chatbot Question Scripts}} & \multicolumn{1}{c}{\textbf{AQ-27 Survey Items}} & \multicolumn{1}{c}{\textbf{Attribution Types}}\\
\midrule
Just like what is mentioned in the story, Avery is currently facing difficulties in both their relationships with colleagues and their work performance. Do you believe Avery's current situation is primarily \textbf{a result of their actions}? Please share your thoughts. & I would think that it was Harry's \textbf{own fault} that he is in the present condition. & Responsibility \\
\specialrule{0em}{1pt}{1pt}
If you were in the process of selecting a \textbf{tenant} for your home, would you feel comfortable entrusting it to someone like Avery? & If I were a \textbf{landlord}, I probably would rent an apartment to Harry. & Social Distance \\
\specialrule{0em}{1pt}{1pt}
Let's imagine you were one of Avery's friends at the party or other social occasions. Would you feel \textbf{anger} toward them due to their actions, for example, losing their temper and yelling at someone? & How \textbf{angry} would you feel at Harry? & Anger \\
\specialrule{0em}{1pt}{1pt}
I wonder if you were one of Avery's colleagues or friends, then would you be inclined to \textbf{help} them with tasks, such as work projects? & How likely is it that you would \textbf{help} Harry? & Helping \\
\specialrule{0em}{1pt}{1pt}
I'd like to know if you would feel \textbf{concern} and \textbf{sympathy} at Avery's mental state? & How much \textbf{sympathy} and \textbf{concern} would you feel for Harry? & Pity \\
\specialrule{0em}{1pt}{1pt}
Apart from being a colleague or friend, I'm also curious if you are Avery's neighbor, then would you consider it better to \textbf{separate} them from the neighborhood and have them undergo \textbf{hospitalization} for treatment? & How much do you think an \textbf{asylum}, where Harry can be \textbf{kept away from} his neighbors, is the best place for him? & Coercive Segregation \\
\specialrule{0em}{1pt}{1pt}
If you were traveling or engaging in activities with Avery, would you have any feelings of \textbf{frightened} or being \textbf{threatened} regarding the potential implications of your interaction? & I would feel \textbf{frightened} and \textbf{threatened} by Harry. & Fear \\
\bottomrule
\end{tabular}
\end{table*}

\subsubsection{Human Coding}
\label{sec:humancoding}

While AI shows promise in qualitative coding, it is crucial to have humans validate its reliability, as relying solely on AI may lead to biases and overprediction \cite{bias_labelling_llm_ashwin_2023}. 
Thus, we conducted human \textbf{deductive coding} \cite{coding_manual_saldana_2016} to establish expert codes for AI-assisted coding, create a training dataset, and develop a codebook that could guide the research team when giving AI instructions.
We formed a coding team of two members as a triangulation approach \cite{coding_manual_saldana_2016}: one author with a computer-science background and a hired graduate student with social-science expertise.
In line with our interview questions and the attribution model \cite{attribution_model_corrigan_2003}, the coding scheme comprised seven stigma attributions (i.e., \textit{responsibility}, \textit{anger}, \textit{pity}, \textit{fear}, \textit{helping}, \textit{coercive segregation}, and \textit{social distance}) and one code for \textit{non-stigmatization}.
The two coders coded messages from 600 randomly selected participants (a total of 4,200 messages) under the guidance of the main researcher and the mental-health specialist, using Cohen's $\kappa$ as a measure of inter-rater reliability \cite{cohens_kappa_mchugh_2012}.

We developed a preliminary codebook and iteratively revised it through several rounds of coding and discussion.
The two coders independently coded messages with frequent agreement checks. 
Starting with shorter intervals, they first coded 10 participants ($\kappa=0.55$), then another 10 ($\kappa=0.53$), followed by four sets of 20 participants each ($\kappa=0.66$, $0.79$, $0.76$, and $0.72$). 
As consistency improved, they extended the intervals to two sets of 50 participants ($\kappa=0.74$, $0.66$), 100 participants ($\kappa=0.69$), and two sets of 150 participants ($\kappa=0.69$, $0.66$). 
At each checkpoint, the two coders engaged in open discussions with the main researcher and the mental-health specialist about their coding decisions until consensus about the coding rules/specifications was reached.
We achieved \textbf{Cohen's $\kappa$ of 0.71} across all 4,200 messages, reaching a satisfactory level \cite{cohens_kappa_mchugh_2012}; and to further validate the coding quality, we randomly sampled and re-coded 50 messages from the first 700 coded messages, 40 from the next 1,400, and 10 from the last 2,100, which yielded a $\kappa$ of 0.69.

\subsubsection{AI-assisted Coding}
\label{method:auto}

To automatically assign labels indicating the presence of specific stigma attributions in each participant's messages, we employed \texttt{GPT-4-Turbo}. 
The prompt-based method we used to transform our human-developed codebook into a set of instructions for the LLM's qualitative coding, as well as the settings and metrics for the validation tests of this coding, are described below.

\paragraph{\textcolor{darkred}{\textbf{Prompt Curation.}}}
To be effective, our instructions to LLMs had to contain several key elements. 
For \textit{context}, these included the vignette, chatbot questions, and participant messages. 
We also required the LLM to select the most appropriate code from a set of predefined options based on the \textit{constraints} it was given, which comprised information about all codes, including their definitions, keywords, coding rules/specifications derived from our human-developed codebook, and selected examples.

We then determined the structure of the prompts based on the results of our pilot tests and prior guidelines \cite{prompt_guidance_ziems_2024, chatbot_various_task_amin_2023, label_llm_kuzman_2023, label_llm_zhu_2023}. 
We used concise language starting with action verbs and employed a multiple-choice format, transforming the coding task into a single-choice question with a set of lettered answer options. 
\textit{Constraints} were provided before the \textit{context} to promote instruction-following, and the output format was regulated. 
We also asked the LLM to provide explanations for its coding decisions and framed our questions as \textit{what-is-your-prediction} instead of \textit{can-you-predict} \cite{chatbot_various_task_amin_2023}. 
Additionally, we utilized role-playing strategies \cite{prompt_practice_reiss_2023}, casting the LLM as \textit{a competent coder for depression stigma}. 
To ensure reproducibility, we set the temperature to zero and generated five outputs for each message, using a majority-voting mechanism to determine the final code. 
The full prompt text is available in the \textit{Supplementary Materials}.

\paragraph{\textcolor{darkred}{\textbf{Validation Settings.}}}
To validate the AI-assisted coding process, we first assessed \textbf{AI-human agreement} across 4,200 messages with existing human codes. 
We then randomly selected 25 messages per code (a total of 200 messages) from the 2,814 AI-coded messages that had not previously been human-coded and proceeded to code them, allowing us to assess AI-human agreement on previously unseen data.

We also established baselines by fine-tuning two \textbf{computational models} commonly used in social-media analysis: \texttt{RoBERTa-base} \cite{roberta_liu_2019} and \texttt{BERTweet-base} \cite{bertweet_nguyen_2020}. 
The 4,200 human-coded message-code pairs $(m, c)$ were shuffled and split into 80\% training and 20\% test sets using stratified sampling to ensure the code distribution in both sets matched that of the complete dataset. 
We compared the agreement (Cohen's $\kappa$) between human coding and each approach (AI-assisted coding and both baseline models) on the test set.
For baseline models, we identified the optimal configuration through hyperparameter tuning: a training regime of 3 epochs (iterations) with batch sizes of 12 and 5 for \texttt{RoBERTa-base} and \texttt{BERTweet-base} respectively, along with optimized settings for learning rate, AdamW optimizer, and early-stopping criteria.

\subsection{Causal Knowledge Graph Construction} 
\label{method:ckg}

\begin{figure*}
    \centering
    \includegraphics[width=0.95\linewidth]{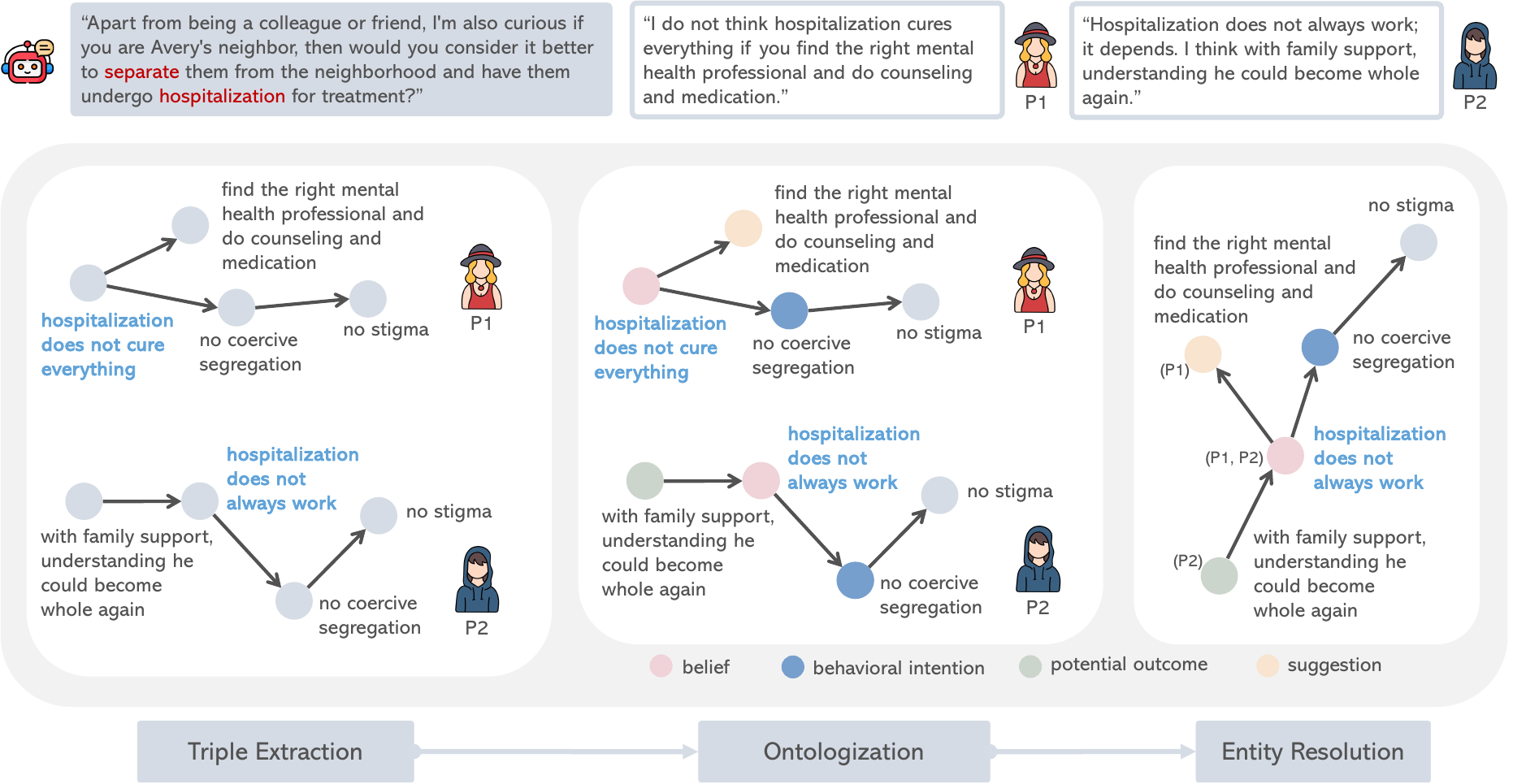}
    \caption{Key steps in the construction workflow for a causal knowledge graph of depression stigma: \textit{Triple Extraction}, where we extract entity-relation-entity triplets from participant messages; \textit{Ontologization}, where we map entities to theoretical constructs; and \textit{Entity Resolution}, where we merge semantically similar entities. These steps lay the foundation for \textit{Conceptual-model Construction}, where we discover emerging themes and interrelationships between constructs (not shown in the figure).}
    \Description{Flowchart showing the transformation of interview responses about depression into a structured causal knowledge graph. Starting with chatbot-collected messages in which participants discuss their views on depression and mental health, the workflow visualizes three key steps: (1) Triple Extraction, where messages are broken down into an entity-relationship-entity format; (2) Ontologization, where extracted entities are categorized into color-coded constructs (pink for beliefs, blue for behavioral intentions, gray for potential outcomes, and beige for suggestions); and (3) Entity Resolution, where semantically similar concepts are merged. The figure sets up for, but does not show, the final Conceptual-model Construction step, where emerging themes and relationships between constructs are identified.}
    \label{fig:graph}
\end{figure*}

We constructed a CKG using messages collected from all 1,002 participants through our chatbot interviews, along with the results of AI-assisted coding (Figure \ref{fig:graph}). 
By organizing this rich interview data into a graph structure, we aimed to illustrate the mechanisms and causalities underlying stigmatizing behavioral intentions at the macro level. 
The relevant workflow consisted of four parts – triple extraction, ontologization, entity resolution, and conceptual-model construction – each of which is explained in detail below. 

For clarity, we distinguish between three key terms in our paper: \textit{\textbf{entities}} (individual nodes in the CKG representing specific text segments), \textbf{\textit{constructs}} (theoretical categories to which \textit{entities} are mapped), and \textbf{\textit{themes}} (recurring patterns identified within each \textit{construct}).

\subsubsection{Triple Extraction}

To extract causal relationships, we conducted prompt-based fine-tuning using \texttt{GPT-3.5-Turbo} \cite{kg_carta_2023}. 
Specifically, for each message, we pre-specified a triple connecting stigmatization status (\textit{stigma} or \textit{no stigma}) to the corresponding codes obtained from human/AI coding. 
For example, a message classified as lacking pity would yield the triple \textit{"(stigma, because, no pity)"}. 
Separately, we fine-tuned the model to extract further causal relationships, such as the reasons behind the lack of pity.

To create annotated data, we began by manually extracting triples from 70 messages – comprising five stigmatizing and five non-stigmatizing responses to each of the seven attribution questions. 
We then fine-tuned the model and used it to generate triples for another 70 (random, unseen) messages. 
Two of the authors reviewed, curated, and corrected the model's predictions, and the corrected triples were used to refine the model further. 
By repeating this process for six iterations, with accuracy improving from 0.47 to 0.66, 0.86, 0.90, and 0.93\footnote{We used accuracy instead of Cohen's $\kappa$ because human curation could add or remove triples that the model missed or incorrectly extracted, making the total number of triples inconsistent between human and model outputs. The accuracy was calculated by treating each triple (entity, relationship, entity) as the basic unit of comparison and computing the ratio of matching triples to the total number of triples.}, we covered about 5\% of the dataset. 
During curation, we categorized model errors into five types: cause-effect reversal, logical inconsistencies, wording inaccuracies (incomplete text in entities), redundancies, and omissions. 
We also steered the model to consider longer causal chains.

After curating about 5\% of the messages, we evaluated the model's performance without further fine-tuning by having it extract triples from 70 (new, unseen) messages and then manually assessing these extractions, achieving \textbf{an accuracy of 0.93}.
The final model was then used to process all 7,014 messages based on its knowledge of the 420 curated messages.

\subsubsection{Ontologization}
\label{sec:ontology}

In knowledge engineering, an ontologization step is crucial to mapping extracted entities into theoretical constructs \cite{ke_iqbal_2013}.

\paragraph{\textcolor{darkred}{\textbf{Conceptualization.}}}

We began the ontologization with conceptualization, adopting \textbf{a combination of deductive and inductive coding approaches}. 
Two authors first independently examined 50 entities, each extracted from a different message to maximize data coverage. 
We established an initial coding scheme based on Corrigan et al.'s attribution theory, comprising four theory-driven constructs \cite{hybrid_fereday_2006}: \textit{signaling event}, \textit{cognitive mediator}, \textit{affective response}, and \textit{behavioral reaction} \cite{attribution_theory_corrigan_2000}. 
For each entity, two authors analyzed both the entity itself and the participant's full message as context. 
When encountering entities that did not fit within these predefined theoretical constructs, the two authors held regular meetings to conceptualize and develop new constructs, discussing their role in driving stigma. 
These newly identified constructs were then incorporated into the coding scheme.

This iterative process proceeded with 50 new entities each time until no new constructs emerged, which occurred after a total of 200 entities had been examined, achieving a \textbf{Cohen's $\kappa$ of 0.78}, indicating an acceptable level of agreement \cite{cohens_kappa_mchugh_2012}.
Two authors then consulted with the mental-health specialist to align the construct definition with theories (if any), refining terms such as "\textit{intention}" to "\textit{motivation}" to reflect motivational orientation \cite{motivational_reviewer_kvaale_2016}, and "\textit{nature}" to "\textit{personality}" \cite{personality_reviewer_steiger_2022}.
We present the full set of constructs in the next section, distinguishing between theory-driven and data-driven constructs.

\paragraph{\textcolor{darkred}{\textbf{LLM-assisted Construct Assignment.}}}

We then assigned these constructs to all extracted entities via prompt learning with \texttt{Claude -3}\texttt{-Opus}\footnote{Based on our pilot tests where Claude outperformed OpenAI GPT models, we used it for both ontologization and entity resolution tasks in this study.}. 
More specifically, we created LLM instructions that included all identified constructs and their definitions. 
Each construct was illustrated with examples, such as "\textit{easygoing}" for the construct "\textit{personality}". 
The LLM was instructed to output both the assigned construct and a concise justification (< 20 words). 
We evaluated human-LLM agreement on our human-coded 200 entities and obtained a \textbf{Cohen's $\kappa$ of 0.77}.

\subsubsection{Entity Resolution}

To handle variations in phrasing and consolidate semantically similar entities, we performed entity resolution.

\paragraph{\textcolor{darkred}{\textbf{Identifying Semantically Similar Entities.}}}
The first step was to identify a set of semantically similar entities for each entity in our dataset. 
To calculate semantic similarity, we represented each entity as a vector using word embeddings.
We used multiple embedding methods\footnote{The embedding methods used include GLoVe, BERT, RoBERTa, DistilBERT, ALBERT, XLNet, S-MPNet, S-GTR-T5, S-DistilRoBERTa, and S-Minilm.} to broaden the semantic coverage and potentially reduce biases \cite{embedding_bias_swinger_2019} that may be present in any single embedding method. 
For each entity, we used each embedding method to find the 10 other entities whose semantic meanings were most similar according to their vector representations\footnote{The value of 10 was chosen after experimenting with values of 1, 5, 10, and 20 \cite{entity_matching_2023}, balancing computational resources with coverage of potentially mergeable entities - larger values could identify more potential matches, but they increased computational overhead, while smaller values risked missing valid matches.}. 
We then combined all the similar entities found by different embedding methods into a single candidate set per entity.

\paragraph{\textcolor{darkred}{\textbf{LLM-assisted Entity Matching.}}}
In the second step, we evaluated which entities within these semantically similar sets should be merged. 
For each entity, we filtered its set of semantically similar entities by retaining only those mapped to the same theoretical construct.
For instance, even if two entities were semantically similar, they would not be considered for merging if one was conceptualized as a \textit{cognitive judgment} and the other as a \textit{signaling event}.
After filtering, each entity had an average of 6.74 potential matches ($SD$ = 33.08).

We prompted \texttt{Claude-3-Opus} to determine whether each remaining entity pair should be merged \cite{entity_matching_2023}. 
To assess the performance, we randomly selected 50 pairs that the LLM had deemed mergeable for evaluation by the main researcher. 
This yielded a \textbf{Cohen's $\kappa$ of 0.90} between LLM and human decisions on whether to merge entities, which indicates almost-perfect agreement \cite{cohens_kappa_mchugh_2012}.

\subsubsection{Conceptual-model Construction}
\label{sec:conceptual}

As a final analytic step after building CKG, our aims were to uncover themes within each construct and to map interrelationships between constructs.

\paragraph{\textcolor{darkred}{\textbf{Identifying Themes within Constructs.}}}
To achieve this, we first applied BERTopic \cite{bertopic_grootendorst_2022}, an advanced topic-modeling technique that leverages BERT-based deep-learning models, to each construct separately. 
This technique generated between 60-100 topics per construct, with the optimal number determined by coherence scores \cite{topic_modeling_practice_liu_2024}\footnote{The technical implementation involved BERT for text embedding, UMAP for dimensionality reduction, HDBSCAN for clustering, and TF-IDF for topic extraction.}.

We then interpreted the topic-modeling results using an open-coding approach. 
After closely reading and familiarizing themselves with the topics generated by BERTopic, two authors independently identified a first-level code for each topic from 1) its top 10 keywords and 2) 30 of its representative messages selected at random \cite{topic_modeling_practice_liu_2024}.
Next, through collective discussion with the mental-health specialist, disagreement resolution, and iterative refinement, these first-level codes were aggregated into overarching \textbf{themes} that captured the key patterns within each construct\footnote{We should clarify that this study consists of three "\textit{coding}" phases: 1) \textit{deductive} coding to identify stigma attributions in messages based on the attribution model \cite{attribution_model_corrigan_2003} (Section \ref{sec:humancoding} and Section \ref{method:auto}), 2) \textit{hybrid deductive-inductive} coding to map entities to theoretical constructs guided by attribution theory \cite{attribution_theory_corrigan_2000} (Section \ref{sec:ontology}), and 3) \textit{inductive} coding to identify themes within each construct (Section \ref{sec:conceptual}).}.

\paragraph{\textcolor{darkred}{\textbf{Distilling Key Interrelationships between Constructs.}}}
To further illustrate key interrelationships between constructs, we developed rules and path-selection algorithms using a heuristic approach based on our qualitative data. 
We considered each relationship as a directed \textit{edge} between constructs. 
The \textit{weight} of each edge was determined by its frequency - specifically, how many participant messages contained that particular relationship. 
We first calculated a threshold for each construct by averaging the weights of all its outgoing edges (total weight sum divided by edge count). 
Next, we retained only those edges whose weights exceeded their respective construct's threshold, thus distilling the most critical relationships in our massive graph dataset.

\section{Results}

\begin{figure*}
    \centering
    \includegraphics[width=0.85\linewidth]{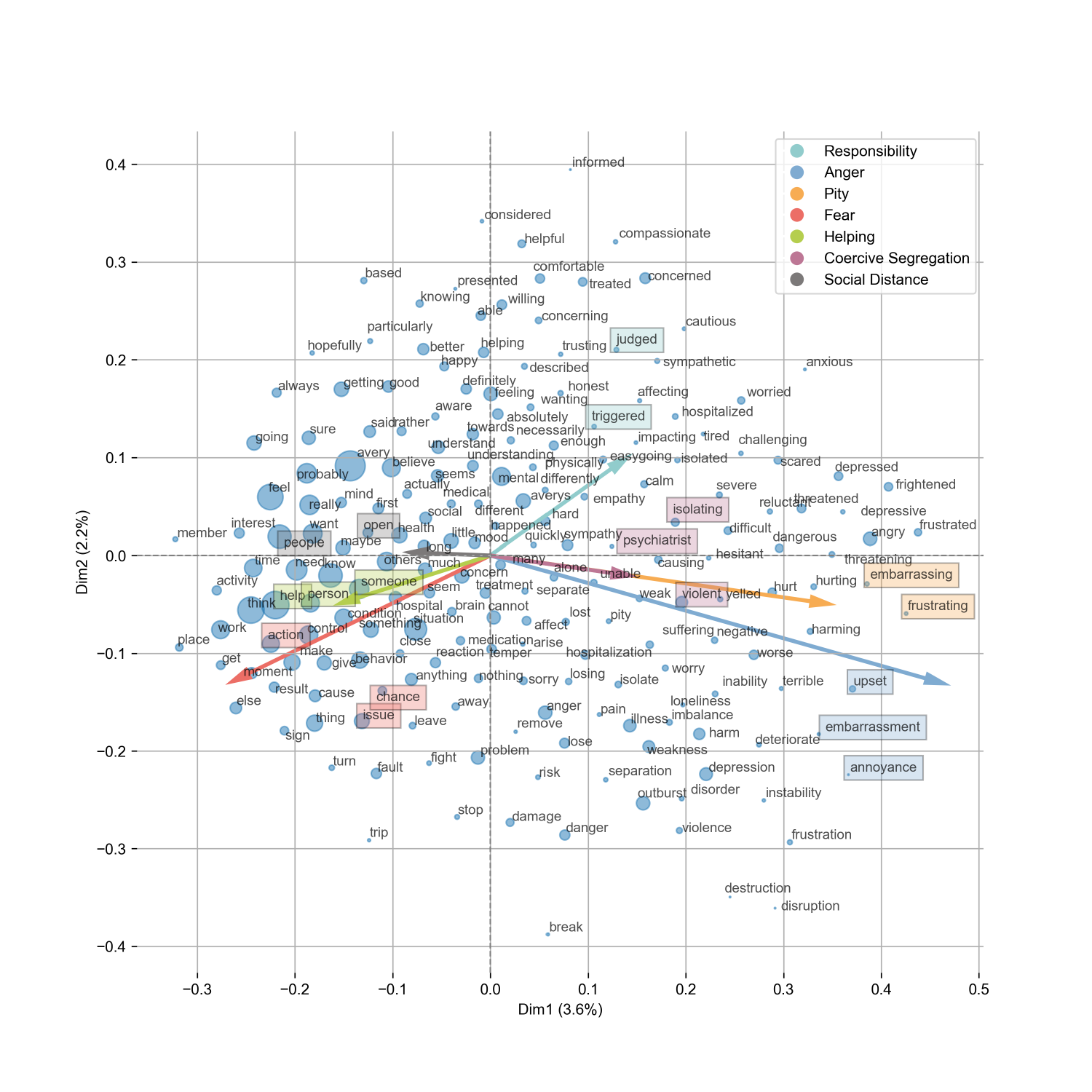}
    \caption{Two-dimensional principal component analysis (PCA) projection of word embeddings from participant messages. The words shown are the most frequent from the 200 $k$-means clusters, and circle sizes represent cluster frequencies. Colored arrows indicate weighted average vectors for different attributions, and word positioning reflects semantic similarity. Highlighted words near attribution arrows represent key terms closely associated with each stigma attribution.}
    \Description{Scatterplot showing a two-dimensional principal component analysis projection of word embeddings from text messages. In this plot, frequently occurring words appear as circles of different sizes, with larger circles indicating higher frequency from the 200 k-means clusters. Seven colored directional arrows cross the plot, each representing different stigma attributions: 'Responsibility,' 'Anger,' 'Pity,' 'Fear,' 'Helping,' 'Coercive Segregation,' and 'Social Distance.' The words are arranged so that semantically similar terms appear closer together. Near each attribution arrow, key related terms are highlighted with boxes.}
    \label{fig:datacollection}
\end{figure*}

\subsection{Assessing AI-assisted Qualitative-data Collection and Analysis Methods for Capturing Depression Stigma (RQ1)}

To address RQ1, we examined two aspects of our AI-assisted pipeline: first, we evaluated the stigma-probing interview data collected by our AI-powered chatbot in terms of 1) data quality, and 2) participant experience with the interview; second, we evaluated the consistency between AI-assisted coding and human coding, while also comparing the agreement among our AI-assisted coding, other computational analytical approaches, and human-derived codes.

\subsubsection{Evaluating AI-assisted Collection of Stigma-related Data}


\paragraph{\textcolor{darkred}{\textbf{Data-quality Evaluation.}}}
We delineated our data along four dimensions based on Gricean Maxims, principles that guide effective communication: the maxim of quantity (\textbf{specificity}), quality (\textbf{self-disclosure}), relation (\textbf{relevance}), and manner (\textbf{clarity}) \cite{metric_grice_1975, chatbot_xiao_2020}. 
Specificity means that participants' responses are detailed and informative; self-disclosure indicates the authenticity and personal nature of the responses; relevance implies that the responses actually address the questions; and clarity facilitates accurate interpretation \cite{metric_grice_1975}.


We first measured the duration of chatbot interviews and the word counts of the participants' contributions to them (two indicators of specificity). 
The average of the former was $t=17.63$ minutes ($SD = 8.18$). 
The mean word counts per participant message, broken down by attribution, were 43.82 for \textit{responsibility} ($SD = 14.68$), 43.40 for \textit{social distance} ($SD = 15.49$), 41.50 for \textit{helping} ($SD = 14.20$), 40.86 for \textit{anger} ($SD = 13.45$), 40.18 for \textit{coercive segregation} ($SD = 13.70$), 39.74 for \textit{fear} ($SD = 14.14$) and 39.30 for \textit{pity} ($SD = 13.98$).

Participants disclosed themselves in their responses, expressing a range of emotions, behavioral intentions, and cognitive reflections. 
For example, P608 shared:

\begin{quote}
    \textit{Not scared of Avery, just concerned for their well-being. I might be concerned about whether or not they would show up and if they would be prepared, dressed appropriately, and such. I might also take on responsibility for booking things if I were worried they would not get that kind of task done.}
\end{quote}

The frequent use of "\textit{I}" alongside personal projections (e.g., "\textit{I might be concerned}", "\textit{I might also take on responsibility}") underscores the authenticity of the responses and the participants' deep engagement with the topic.


The dimensional plot in Figure \ref{fig:datacollection} reveals words' frequencies and semantic similarities, as well as attribution-specific language patterns: e.g., "\textit{frustrating}," "\textit{embarrassment}," and "\textit{upset}" clustered near the \textit{anger} attribution, and "\textit{separate}" and "\textit{psychiatrist}" associated with \textit{coercive segregation}. 
This clustering of semantically related terms around specific stigma attributions shows clear topical alignment, suggesting their relevance.

\begin{figure}
    \centering
    \includegraphics[width=\linewidth]{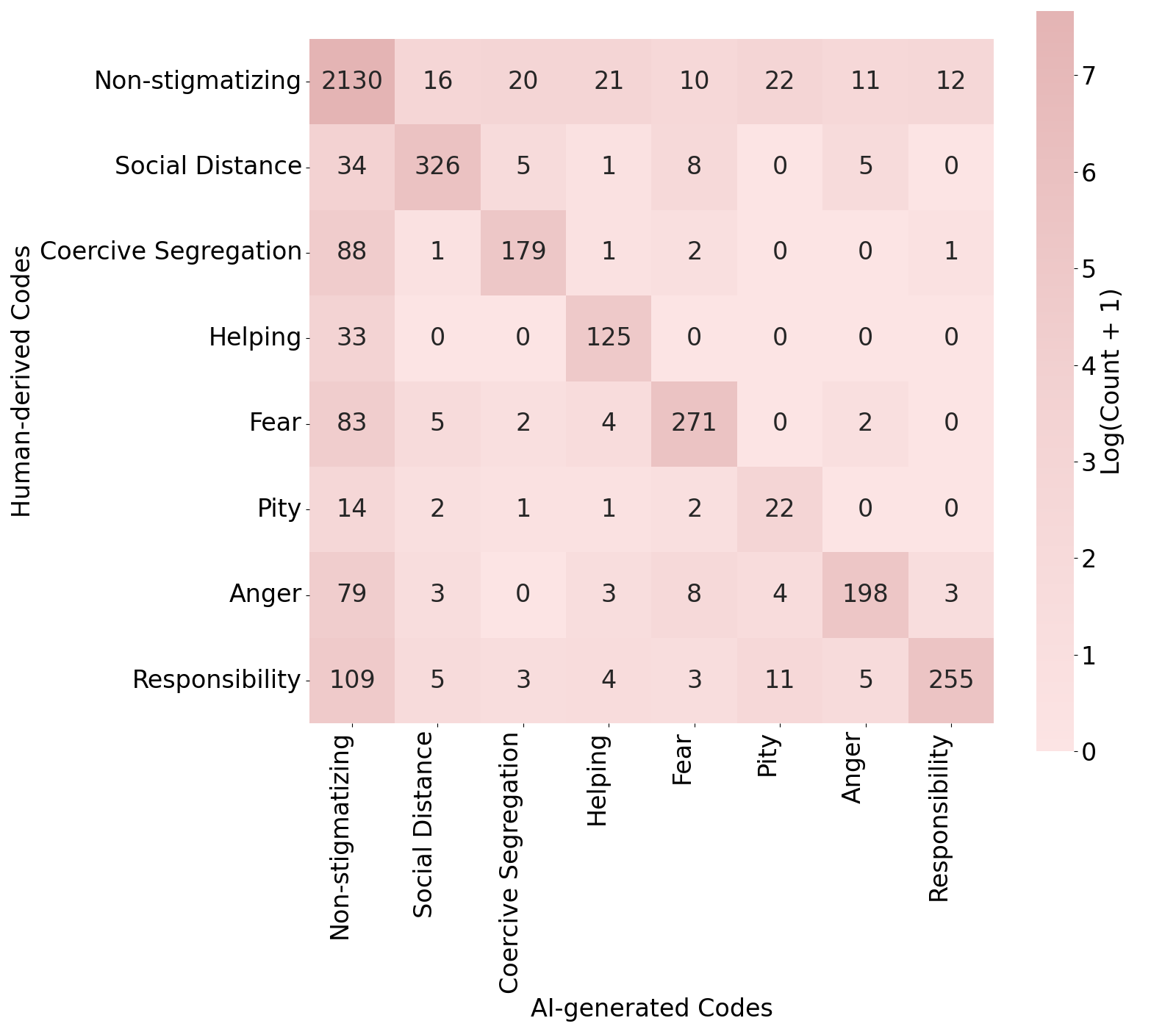}
    \caption{Heatmap showing the agreement between human-derived codes and AI-generated codes. 
    The numbers in each cell represent the frequency of consistency, and darker colors indicate closer agreement.}
    \Description{Heatmap showing the agreement between human-derived and artificial intelligence-generated codes for various stigma-related attributions. The x-axis shows the artificial intelligence-generated codes, while the y-axis shows the human-derived codes. Cell values represent the frequency of agreement, with darker shades indicating higher agreement.}
    \label{fig:heatmap}
\end{figure}

Finally, clarity was assessed by human coding of 4,200 messages from the randomly selected 600 participants. 
Only 1.12\% ($n=47$) were discarded for being incomplete, illegible, off-topic, or excessively brief (<5 words).

\paragraph{\textcolor{darkred}{\textbf{Participant-experience Evaluation.}}}
The 1,002 participants' average response to a five-point Likert-scaled question about their satisfaction with the chatbot interview was 4.37 ($SD = 0.87$), indicating highly positive sentiment toward that experience. 
Among the nearly half of participants who volunteered comments at the end of the chatbot survey ($n=452$), many described it as enjoyable and/or one of their best survey experiences. 
Participants also often highlighted the \textit{thought-provoking} nature of the experience. 
For instance, P607 noted:

\begin{quote}
    \textit{Interacting with the chatbot made me stop and consider my answers, rather than just agree/disagree. It made me examine my actual reactions to a situation when the simple categories did not accurately reflect my concern over what I consider gray areas.}
\end{quote}

\subsubsection{Comparing AI-assisted Coding against Human Coding and Existing Computational Approaches}


\paragraph{\textcolor{darkred}{\textbf{Consistency between AI-assisted Coding and Human Coding.}}}

Of the 4,153 human-coded messages, 46.01\% ($n=1,911$) included stigmatization.
Among the \textit{stigmatizing} codes, the most prevalent was \textit{responsibility}, accounting for 9.51\% ($n=395$). 
\textit{Social distance} was a close second at 9.13\% ($n=379$), while \textit{fear} made up 8.84\% ($n=367$). 
\textit{Anger} and \textit{coercive segregation} constituted 7.18\% ($n=298$) and 6.55\% ($n=272$) of the codes, respectively. 
The least-frequent \textit{stigmatizing} codes were \textit{helping}, at 3.80\% ($n=158$), and \textit{pity}, at 1.01\% ($n=42$).

We calculated Cohen's $\kappa$ to gauge the agreement between human and AI-assisted coding. 
As shown in Figure \ref{fig:heatmap}, we found that it varied across stigma attributions, with \textit{social distance} showing the highest agreement ($\kappa=0.76$), followed by \textit{fear} ($\kappa=0.71$). 
\textit{Responsibility} and \textit{helping} showed similar levels of agreement ($\kappa=0.69$), while \textit{anger} ($\kappa=0.65$), \textit{coercive segregation} ($\kappa=0.54$), and \textit{pity} ($\kappa=0.46$) showed comparatively lower coding consistency.
The overall $\kappa$ across all 4,153 human-coded messages was \textbf{0.69}, indicating a satisfactory level of agreement \cite{cohens_kappa_mchugh_2012} between our AI-assisted and human coding. 
Further, our validation on the 200 previously human-uncoded messages achieved \textbf{Cohen's $\kappa$ = 0.87}, suggesting that our AI-coding approach was not only reliable across the entire dataset but could also be useful for real-time stigma identification given its consistency with human coding.

\begin{figure}
    \centering
    \includegraphics[width=\linewidth]{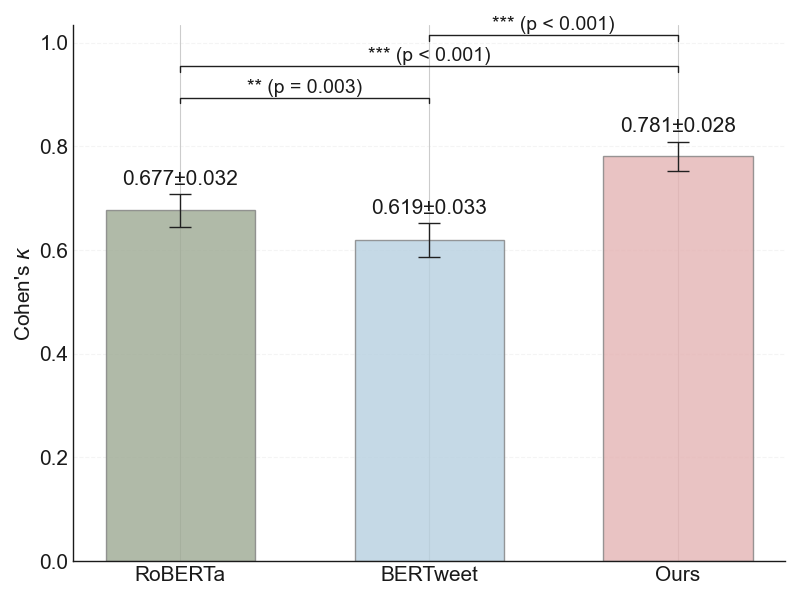}
    \caption{Comparison of human-AI agreement (Cohen's $\kappa$) across different models, showing statistically significantly higher agreement for our AI-assisted coding ($\kappa = 0.781 \pm 0.028$) compared to \texttt{RoBERTa-base} ($\kappa = 0.677 \pm 0.032$) and \texttt{BERTweet-base} ($\kappa = 0.619 \pm 0.033$). Statistical significance levels are marked as p $\geq$ 0.05 (ns), p < 0.05 (*), p < 0.01 (**), or p < 0.001 (***).}
    \Description{Bar chart comparing Cohen's kappa values, which measure the agreement between humans and artificial intelligence, for three different models. The y-axis displays Cohen's kappa scores. The visualization shows three bars: RoBERTa-base with a kappa value of 0.677 plus or minus 0.032, BERTweet-base with a kappa value of 0.619 plus or minus 0.033, and our approach with the highest agreement with a kappa value of 0.781 plus or minus 0.028. The difference between our method and RoBERTa is significant at p < 0.001, as is the difference between our method and BERTweet. The difference between RoBERTa and BERTweet is significant at p = 0.003.}
    \label{fig:computational}
\end{figure}

\paragraph{\textcolor{darkred}{\textbf{Comparison against Existing Computational Methods.}}}

We compared the agreement between human coding, on the one hand, and on the other, 1) our method and 2) the fine-tuned \texttt{RoBERTa-base} \cite{roberta_liu_2019} and \texttt{BERTweet-base} \cite{bertweet_nguyen_2020}, on the \texttt{RoBERTa-base} and \texttt{BERTweet -base} test set of 829 messages (446 human-coded as \textit{non-stigmatizing}, 79 as \textit{responsibility}, 76 as \textit{social distance}, 73 as \textit{fear}, 60 as \textit{anger}, 54 as \textit{coercive segregation}, 32 as \textit{helping}, and 9 as \textit{pity}). 
Our AI-assisted coding yielded higher agreement with human coding ($\kappa = 0.78$) compared to both \texttt{RoBERTa-base} ($\kappa = 0.68$) and \texttt{BERTweet-base} ($\kappa = 0.62$).

A Cochran's Q test revealed significant differences between the models ($\chi^2(2) = 60.24$, $p < .001$) (Figure \ref{fig:computational}). 
Post-hoc analysis using pairwise McNemar tests with Bonferroni correction confirmed significant differences in agreement with human coding between our approach and both \texttt{RoBERTa-base} ($\chi^2(1) = 41.00$, $p < .001$) and \texttt{BERTweet-base} ($\chi^2(1) = 36.00$, $p < .001$).
Between the baseline models, \texttt{RoBERTa-base} showed higher consistency with human coding than \texttt{BERTweet-base} ($\chi^2(1) = 38.00$, $p = .003$).
Detailed significance tests for each code can be found in the \textit{Supplementary Materials}.


\subsection{Integrating CKG with LLM to Uncover Interrelationships between Constructs Underlying Depression Stigma (RQ2)} 

We approached RQ2, regarding the potential of CKG and LLM to synergistically elucidate factors that influence stigmatizing attitudes and discriminatory behaviors, by building CKGs and analyzing the semantics, relationships, and constructs derived from our interview dataset.

\subsubsection{Causal Knowledge Graph Construction}

Our CKG, comprising 13,434 unique entities (reduced from 24,201 entities through entity resolution) interlinked by 18,875 relationships, provides a rich interrelational network of stigma constructs and causalities.

\paragraph{\textcolor{darkred}{\textbf{Constructs Assigned to Entities.}}}

Besides two predefined entities ("\textit{stigma}" and "\textit{no stigma}"), all entities in our dataset were mapped into 11 theoretical constructs.

Specifically, four of these constructs were \textbf{theory-driven}, derived \textbf{deductively} from Corrigan et al.'s attribution theory \cite{attribution_theory_corrigan_2000, theory_overview_corrigan_2002}: \textit{signaling event}, \textit{cognitive judgment}, \textit{emotional response}, and \textit{behavioral intention}.
\textit{Signaling event} ($n$ = 1,365) represented symptomatic behaviors and emotional characteristics of the person with depression depicted in the vignette (e.g., \textit{"Avery feeling judged by others"}, P166), whereas \textit{cognitive judgment} ($n$ = 1,505) captured participants' cognitive evaluation and/or appraisal of that person (\textit{"they are not dangerous"}, P360). 
\textit{Emotional response} ($n$ = 1,017) denoted affective reactions, feelings, and sentiments experienced by participants in response to the vignette (\textit{"feel embarrassed by Avery"}, P357). 
\textit{Behavioral intention} ($n$ = 1,607) indicated participants' anticipated behavioral responses or action tendencies toward the person in the vignette and/or people with depression in general (\textit{"ask them to leave"}, P370).

The remaining seven constructs emerged \textbf{inductively} through the \textbf{data-driven} conceptualization of participant messages: \textit{belief}, \textit{past experience}, \textit{personality}, \textit{situation}, \textit{potential outcome}, \textit{motivation}, and \textit{suggestion}.
\textit{Belief} \cite{belief_reviewer_peter_2021} ($n$ = 2,534) portrayed general knowledge, literacy, attitudes, or deep-seated views about mental health or human elements that participants possess (\textit{"brain is such a complex thing"}, P380). 
\textit{Past experience} \cite{factor_chandra_2007} ($n$ = 782) described participants' prior self-experiences, exposures, and interactions (\textit{"previously had a colleague who had narcissistic tendencies"}, P226). 
\textit{Personality} \cite{personality_reviewer_steiger_2022} ($n$ = 687) depicted participants' self-reported nature, individual traits, or dispositional characteristics (\textit{"I am easy-going"}, P657). 
\textit{Situation} \cite{situation_theory_rusch_2009} ($n$ = 575) delineated participants' immediate extrinsic environmental context (\textit{"I am out at work most of the time"}, P346). 
\textit{Potential outcome} ($n$ = 1,632) involved participants' anticipation of consequences or prognoses for the figure in the vignette (\textit{"Avery will go downhill"}, P325). 
\textit{Motivation} \cite{motivational_reviewer_kvaale_2016} ($n$ = 722) encapsulated participants' drives and what they were striving to achieve or avoid (\textit{"want a tenant who was more reliable"}, P339); and \textit{suggestion} ($n$ = 1,006), their proposed interventions or other recommendations for the figure in the vignette (\textit{"suggest meeting with a professional counselor"}, P524).

\paragraph{\textcolor{darkred}{\textbf{CKG Quality Assessment.}}}

Beyond the subjective evaluations of LLM-human agreement in triple extraction (accuracy = 0.93), ontologization (Cohen's $\kappa$ = 0.77), and entity resolution (Cohen's $\kappa$ = 0.90) reported in Section \ref{method:ckg}, we further assessed our CKG's quality using objective metrics adapted from knowledge-graph-evaluation frameworks \cite{kg_evaluation_koutsiana_2024}\footnote{We excluded the \textit{succinctness} metric, which refers to the CKG's conciseness, as edge repetitions in our context represent meaningful frequency of causal relationships mentioned by participants rather than redundancy.}.

\textbf{Coverage}, which refers to \textit{completeness} and \textit{representativeness}, was first evaluated, showing that our CKG captured much of the participants' reasoning and mental processes.
For \textit{entity} coverage, the text segments we extracted as entities accounted for 61.04\% of all words in the original messages. 
Regarding \textit{relationship} coverage, we captured an average of 3.94 causal relationships per message ($SD$ = 1.52) and 26.34 causations per participant ($SD$ = 6.27). 
For \textit{construct} coverage, each message was associated with an average of 2.83 constructs defined above ($SD$ = 0.75), while individual participants expressed thoughts related to 7.94 constructs on average ($SD$ = 1.32). 
In addition, as discussed in Section \ref{results:conceptual}, our CKG covered and made sense of several established theories of mental-illness stigma.

\begin{figure*}
\centering
\includegraphics[width=0.85\linewidth]{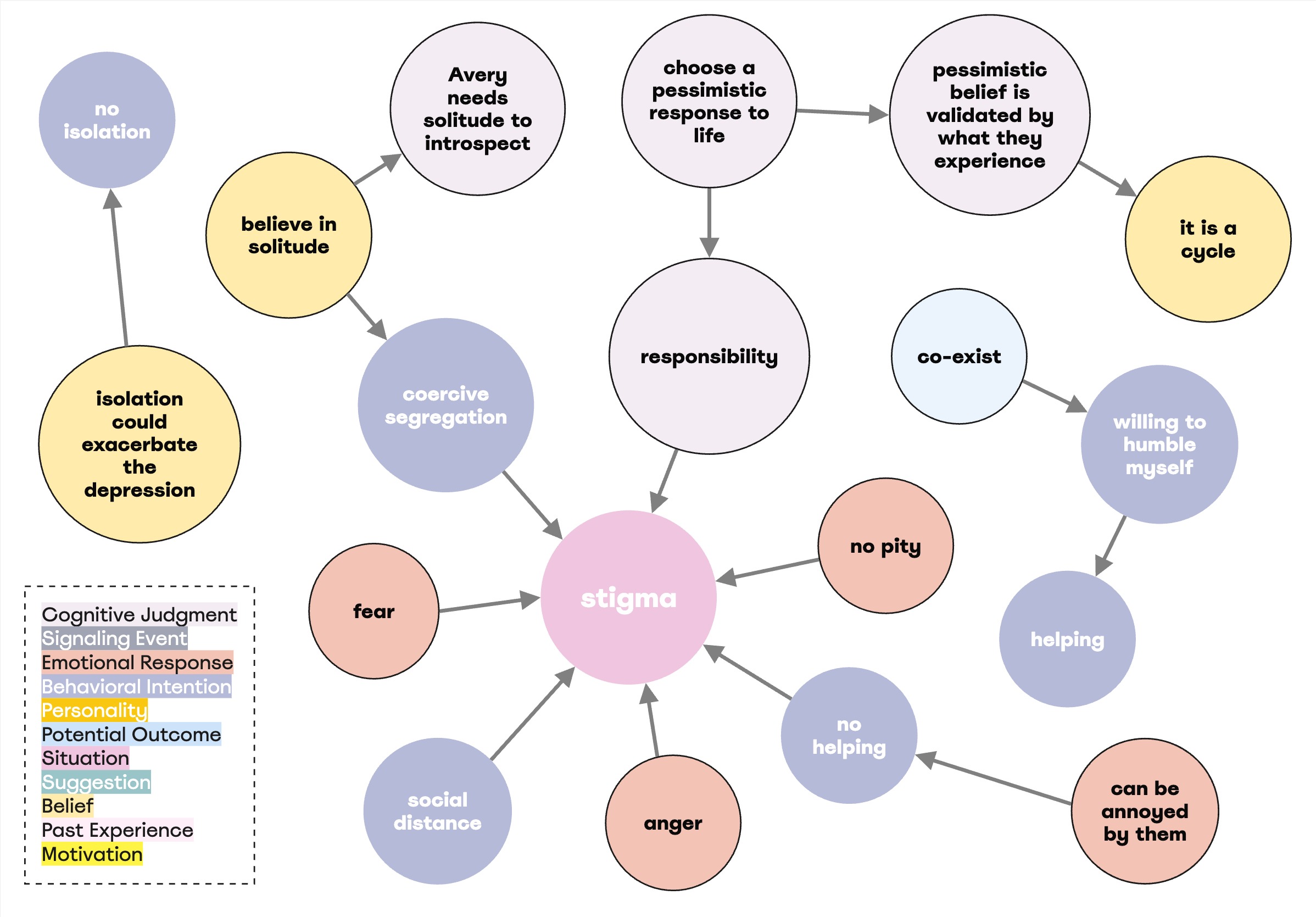}
\caption{Subgraph for P307, who exhibited stigmatizing responses, highlighting the interplay among negative beliefs, emotional responses, and behavioral intentions.}
\Description{Directed graph visualizing participant P307's stigma-related attitudes. Centered on a 'stigma' node, the graph shows interconnected nodes representing different types of responses, e.g., cognitive judgments in white (e.g., 'choose a pessimistic response to life'), beliefs in yellow (e.g., 'believe in loneliness'), emotional responses in coral (e.g., 'anger,' 'fear'), and behavioral intentions in blue (e.g., 'help,' 'social distance'). Nodes are connected by arrows indicating causal relationships, and a legend in the lower left identifies 11 color-coded constructs.}
\label{fig:stigmatizing_subgraph}
\end{figure*}

To evaluate \textbf{coherence}, we examined our CKG through the lens of \textit{consistency} - whether the CKG is free of contradictions and consistent with the domain it represents - and \textit{validity}. 
The analysis of cyclic relationships (as part of validity) revealed only eight cycles in total, traversing a small proportion (1.34\%) of all entities.
We speculated that most of the identified cycles represent reasonable reciprocal causation \cite{reciprocal_mulatu_2002} between constructs.
Connectivity analysis (as part of internal consistency) demonstrated the CKG's strong integration: a single large connected component comprised 98.59\% of the entities, while the remaining 83 small disconnected components collectively comprised only 189 entities.
Further, the causal relationships in our CKG had a mean length of 2.07 steps\footnote{Length here refers to the number of consecutive relationships - for example, (\texttt{entity1}, because, \texttt{entity2}) and (\texttt{entity2}, because, \texttt{entity3}) represent a two-step chain.} ($SD$ = 0.58), demonstrating the CKG's capacity to capture both direct causations and more complex indirect pathways that suggest potential mediating or moderating relationships between constructs.

Finally, we illustrated the potential \textbf{practical utility} of insights derived from our CKG by applying them to decipher individual stigmatizing attitudes, develop conceptual models, and generate hypotheses, which are presented in the next two sections.

\begin{table*}[htbp]
  \centering
  \small
  \caption{Themes and their prevalence rates derived through topic modeling and open coding across 11 stigma-related constructs.}
  \Description{Two-column table showing prevalent themes and their percentage rates related to depression stigma, organized by theme. All themes are listed with their corresponding prevalence rates ranging from 4.10\% to 89.26\%, with each percentage indicating how often that particular theme appeared in participants' messages.}
    \begin{tabular}{p{16.70em}clc}
    \toprule
    \textbf{Theme(s)} & \multicolumn{1}{p{5em}}{\textbf{Prevalence}} & \multicolumn{1}{p{16.70em}}{\textbf{Theme(s)}} & \multicolumn{1}{p{5em}}{\textbf{Prevalence}} \\
    \midrule
    \multicolumn{2}{p{16.70em}}{\textbf{Signaling Event}} & \multicolumn{2}{p{16.70em}}{\textbf{Past Experience}} \\
    \midrule
    Mood Swings & 34.78\% & \multicolumn{1}{p{16.70em}}{Personal Exposure} & 89.26\% \\[3.9pt]
    Social Withdrawal & 27.74\% & \multicolumn{1}{p{16.70em}}{Experiential/Attitudinal Knowledge} & 10.74\% \\
\cmidrule{3-4}    Aggressive, Self-neglect Behaviors & 16.78\% & \multicolumn{2}{p{16.70em}}{\textbf{Personality}} \\
\cmidrule{3-4}    Functional Impairment & 15.26\% & \multicolumn{1}{p{16.70em}}{Sociable, Easygoing, and Even-tempered} & 67.15\% \\[3.9pt]
    Treatment Non-adherence & 5.44\% & \multicolumn{1}{p{16.70em}}{Interpersonally Skeptical, Strong-minded, Stubborn, and Self-reliant} & 32.85\% \\
    \midrule
    \multicolumn{2}{p{16.70em}}{\textbf{Cognitive Judgment}} & \multicolumn{2}{p{16.70em}}{\textbf{Situation}} \\
    \midrule
    Controllability, Personal Responsibility & 42.85\% & \multicolumn{1}{p{16.70em}}{Time Scarcity} & 46.00\% \\[3.9pt]
    Severity, Need for Intervention & 27.02\% & \multicolumn{1}{p{16.70em}}{Physical Attributes} & 28.00\% \\[3.9pt]
    Perceived Dangerousness & 12.29\% & \multicolumn{1}{p{16.70em}}{Time Abundance} & 26.00\% \\
\cmidrule{3-4}    Unpredictability, Instability & 10.83\% & \multicolumn{2}{p{16.70em}}{\textbf{Potential Outcome}} \\
\cmidrule{3-4}    Inability to Judge without Context & 7.01\% & \multicolumn{1}{p{16.70em}}{Self-destructive Tendencies} & 32.88\% \\
\cmidrule{1-2}    \multicolumn{2}{p{16.70em}}{\textbf{Emotional Response}} & \multicolumn{1}{p{16.70em}}{Social Isolation, Relationship Difficulties} & 26.14\% \\
\cmidrule{1-2}    Sympathy, Concern & 65.95\% & \multicolumn{1}{p{16.70em}}{Functional Impairment} & 25.39\% \\[3.9pt]
    Frustration, Irritation & 18.74\% & \multicolumn{1}{p{16.70em}}{Symptom Deterioration} & 15.59\% \\
\cmidrule{3-4}    Fear, Anxiety, and Caution & 11.22\% & \multicolumn{2}{p{16.70em}}{\textbf{Motivation}} \\
\cmidrule{3-4}    Confusion & 4.10\% & \multicolumn{1}{p{16.70em}}{Altruistic, Prosocial} & 72.75\% \\
\cmidrule{1-2}    \multicolumn{2}{p{16.70em}}{\textbf{Behavioral Intention}} & \multicolumn{1}{p{16.70em}}{Egoistic} & 22.63\% \\
\cmidrule{1-2}    Supportive Engagement & 57.19\% & \multicolumn{1}{p{16.70em}}{Reciprocal} & 4.63\% \\
\cmidrule{3-4}    Conflict Management, De-escalation & 16.87\% & \multicolumn{2}{p{16.70em}}{\textbf{Suggestion}} \\
\cmidrule{3-4}    Protective Measures, Avoidance & 15.72\% & \multicolumn{1}{p{16.70em}}{Professional/Authority Help, Treatment} & 65.02\% \\[3.9pt]
    Segregation, Compartmentalization & 10.22\% & \multicolumn{1}{p{16.70em}}{Support System, Social Connections} & 26.77\% \\
\cmidrule{1-2}    \multicolumn{2}{p{16.70em}}{\textbf{Belief}} & \multicolumn{1}{p{16.70em}}{Self-management, Self-improvement} & 8.20\% \\
    \midrule
    Universal Human Experience/Equality & 38.51\% &       &  \\[3.9pt]
    Mental Health Awareness & 35.16\% &       &  \\[3.9pt]
    Social Interconnectedness & 14.65\% &       &  \\[3.9pt]
    Complexity of Innate Human Element & 11.68\% &       &  \\
    \bottomrule
    \end{tabular}%
  \label{tab:theme}%
\end{table*}%

\subsubsection{Case Studies: Participant-specific Subgraphs}

We deconstructed the stigmatizing attitudes of one participant, P307, who invariably exhibited such views, illustrating the feasibility of deconstructing stigma-related patterns in real time.
A parallel example in which we disentangled non-stigmatizing attitudes can be found in the \textit{Supplementary Materials}.

The subgraph for P307 (Figure \ref{fig:stigmatizing_subgraph}) reveals a network of negative beliefs and behaviors, including a causal chain from \textit{"choose a pessimistic response to life"} to \textit{"pessimistic belief is validated by what they experience"} and finally to \textit{"it is a cycle"}. 
This chain of reasoning suggests that the participant attributed the vignette character's mental-health issues to a self-perpetuating cycle of negative choices and experiences, placing \textit{responsibility} on the individual.

The subgraph also depicts the participant's ambivalent feelings about \textit{helping}: annoyance with the vignette's protagonist led to a reluctance to help them. 
Interestingly, however, there was also a path from \textit{"willing to humble myself"} to \textit{"co-exist"}, which resulted in a behavioral intention to help. 
This corresponded to P307's quote, \textit{"nobody is perfect. In as much as I can be annoyed by them I am willing to humble myself enough to co-exist."} 
Thus, while P307 is ostensibly expressing an offer of support, the causal analysis reveals a veiled \textit{microaggression}.

P307 also took a somewhat paradoxical stance on \textit{coercive segregation}, as evidenced by the co-occurrence of two \textit{beliefs} that appear to be contradictory: \textit{"believe in solitude"} and \textit{"isolation could exacerbate the depression"}. 
This was derived from the participant's statement, \textit{"I do believe in solitude but not in isolation. Avery needs solitude to introspect; however, isolation could exacerbate their depression."} 
So, while P307 attempted to differentiate between \textit{solitude} and \textit{isolation}, their reasoning still placed the onus on Avery to engage in introspection, and implied Avery's need for a space away from other people in which to do this. 
This causal co-action subtly reinforced the idea of hospitalization or segregation, even as the participant tried to soften it by acknowledging its potential negative outcomes. 
This reveals how stigmatizing attitudes can be rooted in seemingly benign intentions; and the graph highlights how stigmatizing beliefs can cascade, reinforce one another, and finally lead to discriminatory actions.

\subsubsection{Conceptual-model Construction}
\label{results:conceptual}

After the CKG was constructed, we established rules and restructuring principles in consultation with the mental-health specialist to better theorize our 11 constructs.
These stipulated that: 1) only relationships observed in at least one participant message could be included in the conceptual model; 2) \textit{potential outcome}, \textit{cognitive judgment}, and \textit{belief} were consolidated due to their shared focus on cognitive interpretations of mental health and depression, albeit at different levels of specificity; 3) \textit{motivation} and \textit{personality} were combined on the grounds that they all reflect relatively dispositional, enduring, and innate traits; 4) \textit{suggestion} was excluded as it primarily represented an outcome rather than a predictor of stigmatizing behavioral intentions; and 5) \textit{signaling event}, \textit{past experience}, \textit{personality}, and \textit{situation} could not be led to by other constructs, due to their innate or fixed nature, and that \textit{behavioral intention} could not lead to other constructs, as our focus was on its formation. 
Finally, in line with attribution theory \cite{attribution_theory_corrigan_2000}, we reorganized the constructs into four layers: \textit{stimuli}, \textit{cognitive mediator}, \textit{emotional response}, and \textit{behavioral intention}.

Accordingly, we identified themes across 11 constructs (Table \ref{tab:theme}) that shape attitudes toward depression, which informed the development of the conceptual model (Figure \ref{fig:theory_stigma}) illustrating key causal relationships (see the \textit{Supplementary Materials} for full results).
Below, we discussed how this conceptual model aligned with and went beyond existing theories.

\begin{figure*}
\centering
\includegraphics[width=0.9\linewidth]{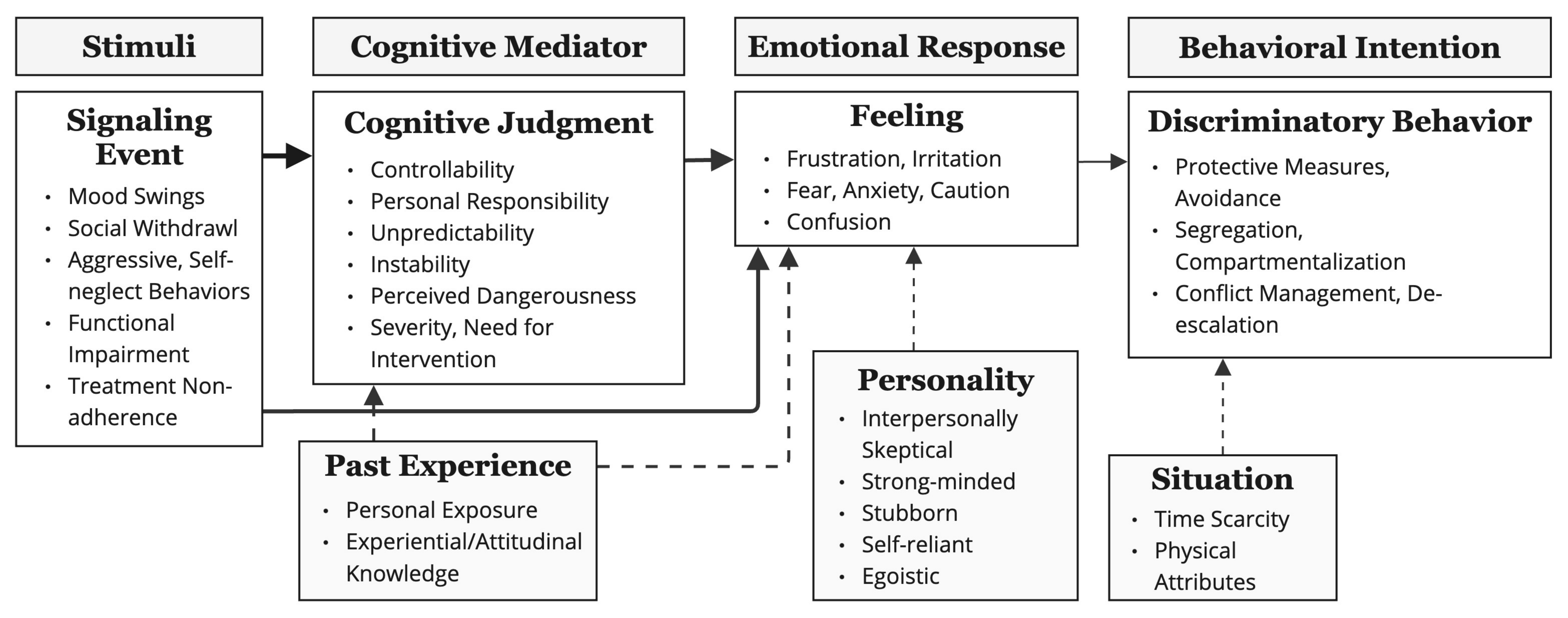}
\caption{Conceptual model of stigmatizing responses derived from the CKG. Solid lines represent pathways known to attribution theory \cite{attribution_theory_corrigan_2000}, while dashed ones indicate new pathways derived from our CKG analysis. The lines' thickness reflects the frequency of those relationships in participant messages.}
\Description{Conceptual model showing how stigmatizing responses develop, organized into four vertical columns: 'Stimuli,' 'Cognitive Mediator,' 'Emotional Response', and 'Behavioral Intention.' Solid lines show pathways known to attribution theory, flowing from 'Signaling Event' to 'Cognitive Judgment' to 'Feeling' to 'Discriminatory Behavior.' Dashed lines indicate new relationships identified through causal knowledge graph analysis, showing how 'Past Experience' influences both 'Cognitive Judgment' and 'Emotional Responses,' 'Personality' impacts 'Emotional Responses,' and 'Situation' directly affects 'Behavioral Intentions.' Line thickness indicates the frequency of the relationship in the data.}
\label{fig:theory_stigma}
\end{figure*}

\paragraph{\textcolor{darkred}{\textbf{Substantiating Existing Theories.}}}

The four theory-driven \textbf{constructs} in our conceptual model - \textit{signaling event}, \textit{cognitive judgment}, \textit{emotional response}, and \textit{behavioral intention} - corresponded directly to those in \textbf{attribution theory} \cite{attribution_theory_corrigan_2000}. 
The specific \textbf{themes} we identified within each construct also mirrored the theory's core findings: our \textit{cognitive-judgment} themes captured assessments of personal responsibility, controllability, unpredictability, and dangerousness; \textit{emotional responses} ranged from sympathy to caution and irritation; and \textit{behavioral intentions} included both supportive engagement and avoidance or compartmentalization.

Our conceptual model confirmed four key \textbf{pathways} consistent with attribution theory \cite{attribution_model_corrigan_2003}: \textit{signaling events} leading to both \textit{cognitive judgments} (extracted from 849 participants) and \textit{emotional responses} (526 participants), as well as the progression from \textit{cognitive judgments} to \textit{emotional responses} (514 participants) and ultimately to discriminatory \textit{behavioral intentions} (65 participants).
The dominance of these paths, particularly the first three, demonstrates that attribution theory's core proposition - that observable symptoms or behaviors of people with mental illness frequently trigger immediate appraisals and emotional responses, which can ultimately lead to discriminatory actions - remains crucial in explaining stigma processes.

Looking beyond attribution theory, several \textbf{other stigma-related theoretical frameworks} help make sense of our data-driven \textbf{constructs} and \textbf{themes}.
The themes under the \textit{belief} construct regarding the universality of personal struggles and the potential for anyone to experience depression align with theoretical work on de-stigmatization that emphasizes continuum rather than categorical beliefs about mental illness, with the former blurring the boundaries between "normal" people and those with psychiatric problems \cite{belief_reviewer_peter_2021}.
Our themes of mental-health awareness (as evidenced by narratives equating mental health with physical health in importance) and the complexity of human nature reflect research exploring how etiological, biological, and psycho-environmental beliefs shape stigma \cite{factor_valery_2020, factor_frost_2024}.


Our \textit{personality}-related constructs and themes, which captured participants' self-descriptions ranging from loyal, gregarious, and people-oriented to reserved, hardheaded, and resolute, align with research showing how openness to experience and agreeableness correlate with reduced discriminatory desires \cite{personality_reviewer_steiger_2022}. 
In the same vein, our \textit{motivational} themes, spanning from self-interest and egoism to prosocial tendencies and benevolence, correspond with theoretical work associating social dominance orientation and the desire to maintain security and social cohesion to mental-illness stigma \cite{motivational_reviewer_kvaale_2016}.

The \textit{situational} themes we identified, including physical attributes and time availability, speak to theories that emphasize how environmental and demographic factors influence stigmatizing attitudes \cite{sct_self_stigma_catalano_2021}; similarly, our findings about personal exposure and experiential knowledge resonate with studies showing how psychosis-like \textit{past experiences} are associated with public stigma in both community and clinical populations \cite{factor_chandra_2007}.
All these alignments substantiate the theoretical validity and value of our CKG-derived findings.

\paragraph{\textcolor{darkred}{\textbf{Novel Pathways beyond Existing Theories.}}}

Uniquely, several \textbf{pathways} unknown to existing theories were also revealed (four dashed lines in Figure \ref{fig:theory_stigma}).
We found that the extrinsic construct \textit{situation} directly influenced \textit{behavioral intentions}, with environmental contexts like time availability modulating pre-existing attitudes and shaping immediate reactions - less-busy participants, for instance, expressed greater willingness to engage with people suffering from depression.
This interaction between situational factors and behavioral responses could be partly made sense of through social-cognitive theory (SCT) \cite{sct_bandura_2001, sct_self_stigma_catalano_2021}, which posited that general human behavior emerges from the interplay of individual characteristics, cognitive factors, and prevailing contexts.
While SCT had once been applied to self-stigma research \cite{sct_self_stigma_catalano_2021}, our findings explored its potential for understanding public stigma, revealing how SCT's situational factors and attribution theory's emotional responses can jointly shape behavioral intentions. 
This intersection suggests new theoretical possibilities for exploring how SCT and attribution theory intertwine in deconstructing public depression stigma.

On the other hand, while prior work had validated intrinsic factors like \textit{personality}'s direct influence on behavioral intentions such as social-distance desires \cite{personality_reviewer_steiger_2022}, our data suggested that personality traits may also shape such desires by modulating \textit{emotional responses} (96 participants). 
For example, we found that self-prioritizing traits were often associated with negative emotions like anxiety or frustration, which could subsequently trigger avoidance tendencies.
This pathway reflected how ingrained dispositional factors served as enduring emotional tendencies, predisposing individuals to experience and express certain emotions more readily when confronted with mental illness-related situations.


And lastly, our data highlighted that \textit{past experiences} informed both \textit{cognitive judgments} (136 participants) and \textit{emotional responses} (74 participants), as individuals drew on their personal history to interpret new situations, often generalizing past interactions with people with depression to guide their current perceptions and feelings. 
Although previous research had suggested that personal experiences might influence self-protective behaviors \cite{experience_weinstein_1989}, our conceptual model generated potential hypotheses that past experiences could also likely shape such stigmatizing responses (e.g., avoidance) by simultaneously affecting cognitive appraisals (e.g., dangerousness) and emotional reactions (e.g., fear).

\subsection{Summary of Results}

In sum, our key findings were:
\begin{itemize}
    \item Our AI-assisted data-collection method effectively captured stigma-related discourses, with participants reporting high satisfaction ($M = 4.37/5$, $SD = 0.87$), engaging in self-disclosure, and providing data with high specificity, relevance, and clarity.
    \item Our AI-assisted coding approach showed high consistency with human-expert coding (Cohen's $\kappa$ = 0.69 overall), and significantly outperformed two existing computational-analysis approaches ($p < .001$).
    \item Our CKG, with 13,434 entities and 18,875 relationships, revealed 11 depression stigma-related constructs and their interrelationships among participants. Based on the CKG, our conceptual model identified themes and pathways that were consistent with existing theories and those that went beyond them.
\end{itemize}

\section{Discussion}

\subsection{AI-assisted Pipeline for Automatic Psychological-construct Decomposition}

Recent HCI studies have highlighted the potential for chatbots to gather fertile data on relatively sensitive topics \cite{chatbot_aq27_practice_lee_2023, disclosure_lee_2022, chatbot_reduce_kim_2020}. 
Our findings support this: we collected over 7,000 messages of approximately 40 words from 1,002 participants discussing mental health through short chatbot interactions averaging less than 20 minutes each.
Likewise, the high consistency we observed between our AI-assisted and human coding resonates with the results of prior studies in which LLM was used to streamline the manual coding of social phenomena \cite{coding_deductive_llm_tai_2024, deductive_labelling_xiao_2023}. 
Our study enriches this body of research by developing \textbf{a unified AI-assisted pipeline} that integrates chatbot-based data collection with AI-assisted coding.

Interestingly, consistency between AI-assisted and human coding varied across stigma attributions, with Cohen's $\kappa$ ranging from as low as 0.46 to as high as 0.76. 
When looking into these human-AI coding divergences, we found that they probably revealed unique analytical perspectives and facilitated a deeper understanding of the socio-psychological constructs being coded.
For example, when dissecting \textit{pity}, the AI identified P437's statement "\textit{I feel emotional and compassionate toward her}" as stigmatizing, reasoning that it could reveal underlying condescension.
Although it lowered coding consistency, this insight challenged the received wisdom, which we initially endorsed, that \textit{pity} is purely positive \cite{attribution_model_corrigan_2003}. 
Such variations may also direct our attention to the underlying differences between human coding - which evolves through collaborative codebook creation, discussions, and oral guidance beyond written rules - and AI coding, which relies on human-written instructions and vast training-data corpora only.
These observations inform future research and highlight the need to systematically examine human-AI coding mismatches, particularly in light of emerging studies of LLMs as potential replacements for human coders \cite{coding_deductive_llm_tai_2024}.


This pipeline was able to both elicit and capture the participants' biased preconceptions and stigmatizing attitudes, even those they might be unconscious of.
We observed entrenched stigma such as \textit{condescension} (\textit{"I can open their eyes and they just need someone to guide them"}, P622), \textit{differential treatment} (\textit{"I will learn what triggers them and walk on eggshells around them"}, P687), and \textit{trivialization} through the dismissal of mental illness as merely a mood issue or overthinking (\textit{"Their problem can be easily overcome"}, P792). 
Uncovering these inadvertent attitudes is vital, given that people who harbor such views may unwittingly act as vectors for the spread and normalization of harmful social stigma. 
We therefore encourage future research to apply our proposed pipeline to nuanced conceptualizations of such microaggressions, microassaults, and microinsults, paving the way for tailored technology-enabled interventions.
In addition, our approach builds upon studies emphasizing language's role in perpetuating stigmatization of and prejudice against marginalized groups \cite{social_media_bail_2017, social_media_pavlova_2020, detect_method_fang_2023, detect_method_mittal_2023, detect_method_roesler_2024}, with a particular focus on the value of chatbot-interview data in understanding psychological constructs such as mental-illness stigma \cite{conversation_importance_jenlink_2005, interview_stigma_measure_lyons_1995}, and as a basis to facilitate deep reflection and personal disclosure. 
It offers new analytical opportunities that complement digital social-media data collection: i.e., less prone to data genericization, difficulty in maintaining context at scale \cite{social_media_decontext_boyd_2012}, proxy-population mismatch, the misidentification of bots as humans, and control of public access to data by proprietary algorithms \cite{social_media_deidentify_ruths_2014}. 
The contextual richness of our data positions our method as a powerful tool that can be expected to be used by scholars in the future to explore psycholinguistic differences between human-chatbot conversations and social media content when investigating psychological constructs.


\subsection{Partnership between LLMs and Causal Knowledge Graphs for Illustrating Psychological Models} 

Our large-scale CKG, featuring more than 13,000 entities and 18,000 relationships, provides a valuable database for exploring psychological mechanisms. 
It allowed us to unpack psychological constructs and build conceptual models that both \textbf{substantiated} core propositions of attribution theory \cite{attribution_theory_corrigan_2000} and other stigma-related frameworks \cite{belief_reviewer_peter_2021, personality_reviewer_steiger_2022, motivational_reviewer_kvaale_2016}, and also \textbf{went beyond them} by generating novel hypotheses such as the dual impact of \textit{past experiences} on both \textit{cognitive} and \textit{emotional} responses.

By allowing psychological constructs to emerge from attained data while being guided by social theories, our approach serves as a bridge between theory-driven methodologies and emerging data-centric paradigms in HCI and psychology \cite{ckg_llm_tong_2024}, revealing mechanisms that may sideline people suffering from mental illness.
Here, however, it should be noted that two points clarify our study's focus.
First, this paper is not to establish robust social-science models; rather, it is to demonstrate our method's potential to extend existing theories and build new ones by rapidly generating multiple hypotheses and uncovering latent pathways. 
Second, we do not work with causal graphs between pairs of variables from quantitative data, as is common in previous LLM-assisted causal reasoning studies \cite{ckg_llm_kiciman_2024, kg_psych_crielaard_2022}; instead, we explore the construction of causal \textit{knowledge} graphs from \textit{qualitative} data, which harbor contextual insights from rich, narrative information.
There are several directions for future work: explore methods to validate causal pathways that LLMs/CKGs unveil from qualitative data; systematically disentangle the concordances and discordances between the conceptual models of stigmatizing and non-stigmatizing responses; and refine our conceptual models by adding factors such as age and gender.

Additionally, our work chimes with recent studies on the synergistic potential of partnering LLMs and CKGs in social-psychological inquiry \cite{ckg_llm_tong_2024, kg_carta_2023, ckg_uleman_2021, ckg_borsboom_2021, kg_llm_pan_2024, kg_psych_crielaard_2022}. 
We have advanced this body of research by integrating \textit{node-level} analysis with the extraction of \textit{overarching} psychological constructs. 
Specifically, we integrated fine-grained LLM semantic analysis and the global perspective on causality provided by CKGs and their conceptual models, mirroring the holistic vs. analytic cognition dichotomy \cite{human_psych_nisbett_2001}. 
Building upon this integration, the resulting large-scale CKG could provide an infrastructure for \textit{link prediction} \cite{ckg_llm_tong_2024}, serving as a springboard for generating novel abductive hypotheses through the discovery of non-obvious or indirect causalities within the intertwined graph structure.
Such methods presumably move beyond \textit{descriptive} analyses like word counting and assigning brief labels to language \cite{theory_nlp_boyd_2021} to provide \textit{explanatory} insights that tease apart the antecedents, facets, and consequences of psychological constructs.
Fruitful avenues for future research include exploring effective retrieval methods for these CKGs; investigating their practical application by social scientists; and examining more complex relationships such as intensifying factors, protective factors, mediation processes, buffering effects, and enhancing effects in broader hyper-relational knowledge-graph contexts.


\subsection{Design Implications}

\paragraph{\textcolor{darkred}{\textbf{Real-time Identification for Tailored Micro-interventions to Combat Stigma}}}



The insights derived from the present study not only provide a foundation for the development of technology-enhanced real-time micro-interventions (e.g., VR-based \cite{realtime_martinez_2024}, game-based \cite{personalized_anvari_2024}) for reducing depression stigma and changing psychological constructs - an important line of HCI research - but also enable \textit{personalization}, which researchers have identified as a critical need in this domain \cite{personalized_anvari_2024}.


Our AI-assisted coding and LLM-CKG approach enable \textit{real-time} identification of stigma attributions and attributing processes. 
This real-time detection allows designers to determine which specific micro-intervention events - highly focused, in-the-moment elements designed to promote emotional, cognitive, or behavioral change \cite{micro_intervention_baumel_2020, micro_intervention_howe_2022} - are most appropriate. 
For example, when detecting responsibility-related beliefs, practitioners might consider deploying targeted didactic materials that teach about the biological and environmental factors \cite{belief_reviewer_peter_2021} of mental illness in the present moment. 
Similarly, when misconceptions about dangerousness arise, the digital system could suggest reframing exercises that tunnel the individual's focus toward alternative perspectives \cite{micro_intervention_howe_2022} to counteract premature negative appraisals.

In addition, our CKG shows the potential to inform therapeutic narratives and intervention delivery \cite{micro_intervention_baumel_2020} through its centralized knowledge structure that captures both theoretical understanding and user-specific context. 
Beyond conceptual models, the built CKG enables the creation of graphs of individuals' unique psychological profiles. 
Intervention designers could leverage these personalized graphs to customize therapeutic-session dialogues using a retrieval-augmented generation approach \cite{rag_chen_2024} that involves querying the CKG for relevant past experiences, belief patterns, and causal attributions.

This automated, individualized approach would allow practitioners to move beyond one-size-fits-all anti-stigma strategies that risk backfire effects \cite{backfire_dobson_2022} to broader context-sensitive micro-interventions aimed at dispelling myths, correcting misconceptions, and reinforcing positive beliefs \cite{realtime_martinez_2024, intervention_corrigan_1999}.



\paragraph{\textcolor{darkred}{\textbf{Cultural Sensitivity of AI-assisted Psychometric Analysis}}}


Our study suggests implications for cultural-sensitive psychometric design. 
Because psychological constructs are deeply influenced by socio-cultural factors \cite{culture_difference_krendl_2020}, the interrelationships captured in our CKG are inherently tied to our participants' cultural backgrounds. 
To adapt our approach to different cultural contexts, the interview protocol and AI-assisted analysis pipeline can be extended to support multilingual processing for multinational and multicultural populations; the CKG-construction methodology could be refined to capture local dialects, colloquial idioms, and vernacular that carry unique psycholinguistic nuances.

CKGs have the potential to elucidate the evolution of psychological constructs across diverse socio-cultural groups, thus enabling cross-cultural analysis of psychological mechanisms \cite{crosscultural_obeid_2015}. 
By applying knowledge-graph alignment techniques \cite{kg_llm_pan_2024} to CKGs from different demographic groups (e.g., Western vs. Eastern societies, different age cohorts), researchers could discern similarities and differences in their conceptual models, revealing culture-specific patterns. 
Examining how our approach to deconstructing psychological constructs translates across different cultural contexts would help validate its global applicability, validity, and generalizability.



\paragraph{\textcolor{darkred}{\textbf{Facilitating Psychological-dataset Creation}}}

Finally, our pioneering combination of chatbots for data collection and high-quality AI-assisted coding is scalable; addresses recent calls for developing datasets that adhere to stringent standards for expert evaluation and impact assessment \cite{llm_psycho_demszky_2023}; and could underpin novel approaches to creating structured, well-coded, varied datasets in psychology and related fields. 
This, in turn, could lay a foundation for more influential theoretical contributions in HCI.

\subsection{Limitations and Future Research}

Several methodological limitations warrant discussion. 
First, because our approach focused on message-level stigma analysis, we could have overlooked dynamics that emerged and evolved in complex ways over the course of the entire interview \cite{conversation_level_paakki_2024}. 
A valuable next step would be to explore how attitudes evolve through interviews, how different chatbot-design strategies affect participant disclosure, and how the interplay between chatbot responses and participant statements shapes the expression of stigmatizing attitudes over time. 
Such temporal analysis of conversation trajectories could reveal important patterns in how participants refine their views through interaction.

Second, there are potentially important differences in the dynamics of human-human and human-chatbot conversations. 
Factors such as non-verbal cues and the ability to pick up on tone subtleties are likely to be limited in chatbot interactions \cite{nonverbal_denham_2013}. 
Future comparative analyses of these two broad types of interaction could enhance the ecological validity of our findings and reveal differences in the disclosure and expression of beliefs, attitudes, and behavioral intentions.

Certain limitations of the present case study should also be acknowledged. 
First, our participant pool was primarily from Western countries, which may have affected our ability to capture the full spectrum of culture-specific stigma-related nuances. 
Future research should prioritize obtaining culturally diverse samples to examine how our approach performs in different socio-cultural contexts. 
Second, given that our human-derived codes are themselves not infallible, more attention and analysis are needed when comparing them with AI-assisted coding.



\section{Conclusion}

This study has introduced a novel approach to computationally deconstructing mental-illness stigma, in which data collection via human-chatbot conversations was synergized with AI-assisted coding. 
Holistic modeling of the focal stigma through a combination of CKGs and LLMs represents a potential new paradigm, and has revealed interrelationships among the factors behind stigmatizing statements that expand on existing theoretical work. 
These results not only deepen our insights into depression stigma, but also have broader methodological and design implications for HCI and psychological research, especially technology-enabled intervention design. 
We call for more research to explore our proposed approach's potential to identify additional patterns and interconnections among psychological factors in real time; to examine socio-cultural variation in stigma formation; and to develop targeted anti-stigma campaigns based on the causal pathways it uncovers.

\begin{acks}

Our research has been supported by the Ministry of Education, Singapore (A-8002610-00-00) and the National University of Singapore Start-up Grant Award (R-124-000-128-133). 
This research work is partially supported by the NUS IT's Research Computing group (NUSREC-HPC-00001) and AWS. 
We are grateful to all the reviewers for their valuable comments and suggestions that helped improve this paper, and to all the participants whose time and efforts made this research possible.
\end{acks}

\bibliographystyle{ACM-Reference-Format}
\bibliography{sample-base}




\end{document}